\documentclass[11pt,a4paper]{article}
\DeclareMathAlphabet{\scr}{U}{rsfs}{m}{n}

\usepackage{cite}
\usepackage[utf8]{inputenc}
\usepackage[T1]{fontenc}
\usepackage[footnotesize]{caption}
\usepackage{subcaption}
\usepackage{amsmath,amssymb}
\usepackage{enumerate}
\usepackage{graphicx}
\usepackage{xcolor}
\usepackage{float}
\usepackage{placeins}               
\usepackage{booktabs}
\usepackage{multirow}
\usepackage{hyperref}
\usepackage{cleveref}               

\usepackage[margin=2.5cm]{geometry}
\usepackage{lscape}                 
\usepackage[nolist,nohyperlinks]{acronym}
\usepackage{siunitx}                

\newcommand{\pt}{\ensuremath{p_T}}
\newcommand{\met}{\ensuremath{\pt^\text{miss}}}
\newcommand{\HT}{$H_T$}
\newcommand{\metphi}{\ensuremath{\phi^\text{miss}}}
\newcommand{\chione}{\ensuremath{\tilde{\chi}^0_1}}
\newcommand{\chionep}{\ensuremath{\tilde{\chi}^+_1}}
\newcommand{\chionem}{\ensuremath{\tilde{\chi}^-_1}}
\newcommand{\chitwo}{\ensuremath{\tilde{\chi}^0_2}}

\newcommand{\se}{\ensuremath{\tilde{e}_{L}}}
\newcommand{\sep}{\ensuremath{\tilde{e}_{L}^+}}
\newcommand{\sem}{\ensuremath{\tilde{e}_{L}^-}}
\newcommand{\nuse}{\ensuremath{\tilde{\nu}_{e}}}
\newcommand{\nusebar}{\ensuremath{\tilde{\nu}_{e}^{*}}}

\newcommand{\smu}{\ensuremath{\tilde{\mu}_{L}}}
\newcommand{\smup}{\ensuremath{\tilde{\mu}_{L}^+}}
\newcommand{\smum}{\ensuremath{\tilde{\mu}_{L}^-}}
\newcommand{\nusmu}{\ensuremath{\tilde{\nu}_{\mu}}}
\newcommand{\nusmubar}{\ensuremath{\tilde{\nu}_{\mu}^{*}}}

\newcommand{\stauone}{\ensuremath{\tilde{\tau}_1}}
\newcommand{\nustau}{\ensuremath{\tilde{\nu}_{\tau}}}

\acrodef{ML}{machine learning}
\acrodef{XAI}{explainable AI}
\acrodef{DNN}{deep neural network}
\acrodef{ROC}{receiver operating characteristics}
\acrodef{AUC}{area under curve}
\acrodef{SM}{Standard Model}
\acrodef{SUSY}{supersymmetry}
\acrodef{LHC}{Large Hadron Collider}
\acrodef{LSP}{lightest supersymmetric particle}
\acrodef{NLSP}{next-to-LSP}
\acrodef{ADL}{attributed dependence on labels}
\acrodef{ADP}{attributed dependence on predictions}

\begin{document}

\title{Beyond Cuts in Small Signal Scenarios\\
\Large Enhanced Sneutrino Detectability Using Machine Learning
}
\date{}
\author{%
Daniel Alvestad$^1$\footnote{\texttt{daniel.alvestad@uis.no}}\;,
Nikolai Fomin$^2$\footnote{\texttt{nikolai.fomin@uib.no}}\;,
Jörn Kersten$^{2,3}$\footnote{\texttt{joern.kersten@uib.no}}\;, 
Steffen Maeland$^4$\footnote{\texttt{steffen.maeland@hvl.no}}\;, 
Inga Str{\"u}mke$^{5,6}$\footnote{\texttt{inga.strumke@ntnu.no}}\\[9mm]
{\small\it $^1$Department of Mathematics and Physics, University of Stavanger, 4021 Stavanger, Norway}\\[1mm]
{\small\it $^2$Department of Physics and Technology, University of Bergen, 5020 Bergen, Norway}\\[1mm]
{\small\it $^3$Korea Institute for Advanced Study, Seoul 02455, Republic of Korea}\\[1mm]
{\small\it $^4$Western Norway University of Applied Sciences, 5063 Bergen, Norway}\\[1mm]
{\small\it $^5$Department of Computer Science, Norwegian University of Science and Technology,} \\
{\small\it 7034 Trondheim, Norway} \\
{\small\it $^6$Department of Holistic Systems, SimulaMet, 0167 Oslo, Norway}\\[1mm]
}

\maketitle

\begin{abstract}
\noindent\textit{%
We investigate enhancing the sensitivity of new physics searches at
the LHC by machine learning in the case of background dominance and a
high degree of overlap between the observables for signal and background.
We use two different models, XGBoost and a deep neural network, to
exploit correlations between observables and compare this approach to
the traditional cut-and-count method.  We consider different methods to
analyze the models' output, finding that a template fit generally
performs better than a simple cut.
By means of a Shapley decomposition, we gain additional insight into
the relationship between event kinematics and the machine learning
model output.
We consider a supersymmetric scenario with a metastable sneutrino as a
concrete example, but the methodology can be applied to a much wider
class of models.
}
\end{abstract}

\thispagestyle{empty}
\vfill
\newpage
\setcounter{page}{1}

\tableofcontents

\bigskip\noindent

\section{Introduction}

The absence of a signal of new particles at the \ac{LHC} may suggest that
new physics is realized in a scenario that is hard to detect due to the
absence or very large mass of new colored particles.  Hence, this study
focuses on setups with dominant electroweak production of color-neutral
new particles and multi-lepton signals from their decays.  The
conventional approach to searches for new physics, also known as
``cut-and-count analysis'', is to apply a set of constraints on
different kinematic variables (called ``cuts'' or ``selection'') that 
improve the signal-to-background ratio.  However, the scenarios we
consider can be challenging for this standard approach due to the small
production cross section and the similarity of signal and background
features.  For such problems, \ac{ML} offers a promising alternative
\cite{Guest:2018yhq,Albertsson:2018maf,Abdughani:2019wuv,Bourilkov:2019yoi,Feickert:2021ajf,Schwartz:2021ftp}.
We investigate how much \ac{ML} can increase the discovery reach, and
whether machine learning models can be trained in such a way that they work in
a large region of parameter space and not just for a single point.  This is an
important issue, in particular in new physics scenarios with many free
parameters, as signal kinematics vary from point to point.

As a concrete example, we consider a \ac{SUSY} scenario with a gravitino
\ac{LSP} whose mass is in the \unit{GeV} range.  In addition, the
\ac{NLSP} is assumed to be a sneutrino $\nustau$, the superpartner of a
left-handed tau neutrino.  Due to the relatively large gravitino mass
and its weak couplings, the sneutrino is stable on time scales relevant
for collider experiments.  The \ac{LHC} phenomenology of this scenario
has been considered before
\cite{Covi:2007xj,Ellis:2008as,Figy:2010hu,Katz:2009qx,Katz:2010xg,Bhattacharya:2011zn,Chala:2017jgg},
but not nearly as extensively as that of the scenarios with a neutralino
\ac{LSP} or a gravitino \ac{LSP} and stau \ac{NLSP}.  Motivated in part
by the nature of the \ac{NLSP}, the presence of two hadronically
decaying taus and one muon is chosen as the signature for signal events.
If the cross sections for production via the strong interaction are small
due to large squark and gluino masses, signatures from electroweak
processes will be crucial for detecting \ac{SUSY}.  Electroweak
\ac{SUSY} processes have a significant \ac{SM} background with very
similar collider signatures.  The task of separating signal from
background is therefore challenging from two angles -- predominance of
\ac{SM} background events in the data and a large overlap between
signal and background characteristics.
As a baseline to be compared to \ac{ML} approaches, we perform simple
cut-and-count analyses estimating the sensitivity of the \ac{LHC}
experiments for two benchmark points in parameter space.  Note that
these analyses are not intended to compete with the level of
sophistication of ATLAS and CMS searches.  As our focus is on \ac{ML}
methodology, an expanded analysis with, for example, more complicated
cuts and a detector simulation would be beyond the scope of the paper
and draw away attention from its main results without affecting the
conclusions.

Machine learning algorithms with the ability to learn non-linear
correlations in high-di\-men\-sion\-al data have already proven useful
in cases with a high degree of overlap between
features.
While gradient-boosted decision trees have been the most commonly used
ML method~\cite{aad2020_evindencetttt,aad2021_whzhprod,CMS_ML}, deep neural networks
have also made their appearance in recent years~\cite{aad2021search,aad2021search_rparity,aad2021_higgsdecay,aad2020_assprod}.
Motivated by this, we investigate how well boosted decision trees and a
tuned deep neural network perform at signal classification, compare the relative performance of
the two, and investigate if they generalize equally well across the
parameter space. We also compare the discovery sensitivity of a simple cut on
the value of the \ac{ML} classifier output with the sensitivity obtained
using mixture estimation based on unbinned template fits of the
classifier outputs. This is novel in the context of SUSY searches.
We show that in many scenarios the template fit
method is beneficial and advocate for its use.

\section{Physics scenario}
\subsection{Parameter space points considered}
\label{sec:scenario}
As a prototype for the type of new physics leading to the signal
considered here, \ac{SUSY} with a gravitino \ac{LSP} and a sneutrino
\ac{NLSP} was chosen.
In order to obtain a mass spectrum where a sneutrino is
lighter than all other superparticles except the gravitino, the soft
mass of at least one of the slepton doublets $\tilde\ell_\text{L}$ has
to be smaller than the soft masses of the superpartners of the
right-handed leptons.  In high-scale scenarios for SUSY breaking, this
situation can be arranged fairly easily for non-universal soft Higgs
masses \cite{Buchmuller:2005ma,Ellis:2008as}.

For the study ten benchmark points with a sneutrino NLSP, a Higgs mass
close to the measured value
\footnote{Given the theoretical uncertainty of the Higgs mass
calculation in the minimal supersymmetric extension of the \ac{SM}~\cite{Slavich:2020zjv}, we consider a deviation of about a
GeV from the measured value acceptable.}
and sufficiently large branching ratios for decays producing
taus are selected.  The superparticle and Higgs mass spectra are
computed by \texttt{SPheno 4.0.3}~\cite{Porod:2003um,Porod:2011nf} and
\texttt{FeynHiggs 2.14.2}~\cite{Bahl:2018qog,Bahl:2017aev,Bahl:2016brp,Hahn:2013ria,Frank:2006yh,Degrassi:2002fi,Heinemeyer:1998np,Heinemeyer:1998yj},
respectively.
\texttt{Herwig 7.1.3}~\cite{Bahr:2008pv,Bellm:2015jjp} serves to calculate the cross
sections for the production of \ac{SUSY} particles while \texttt{SPheno
4.0.3} computes the branching ratios of their decays.  

\begin{figure}
    \centering
    \includegraphics[width=1.0\textwidth]{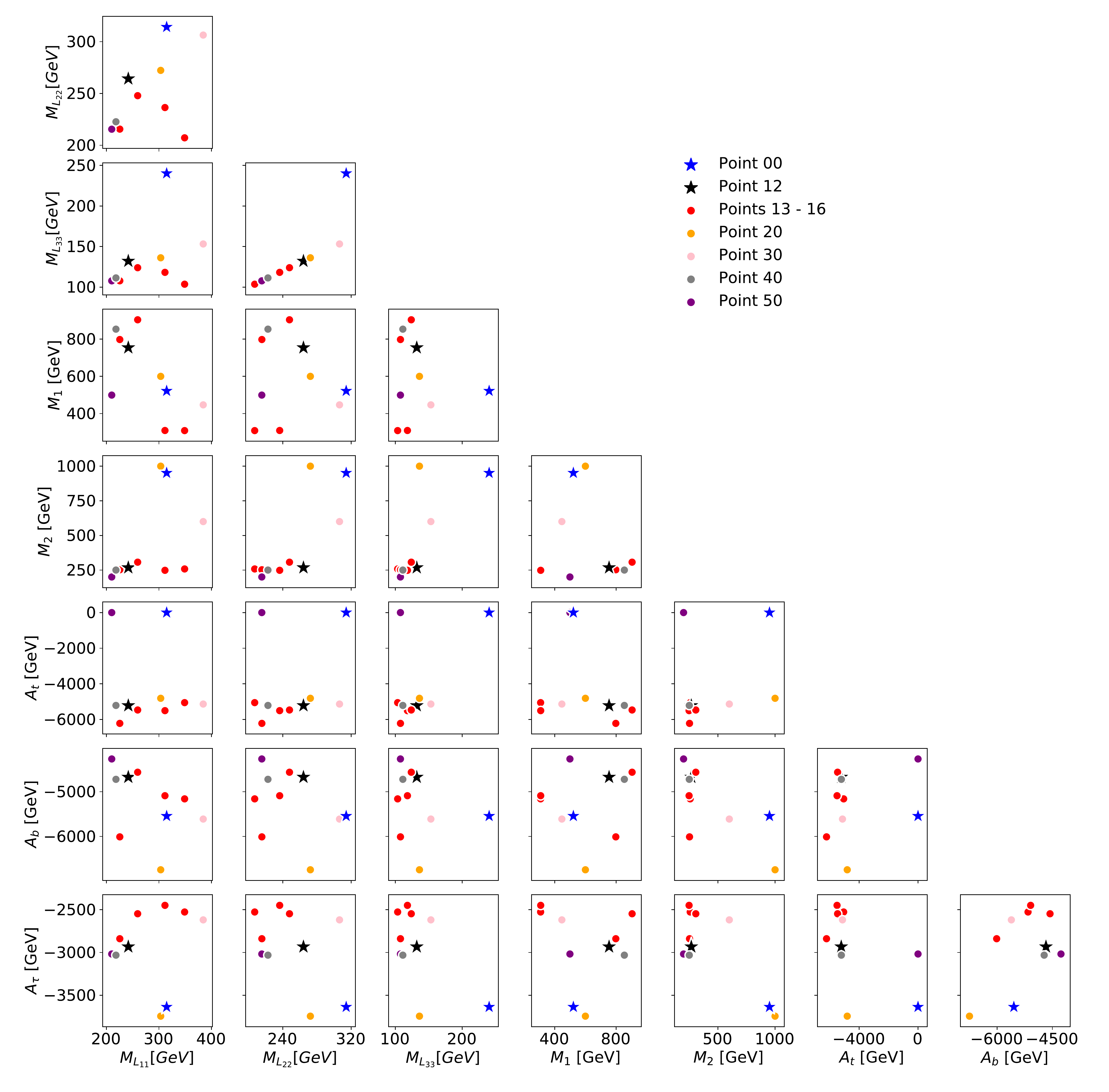}
    \caption{\label{fig:paramspace}Overview of SUSY-breaking parameters
    for the points considered in the analysis. See \cref{app:points} for
    complete details.}
\end{figure}

The benchmark points are shown in \cref{fig:paramspace}.
The detailed input parameters are reported in \cref{app:points}.
These points represent qualitatively different parts of the
parameter space, covering in particular a wide range of $M_2$ and $A_t$
because these parameters have the biggest impact on the signal yield.
As the focus is on the
methodology for new physics searches, no attempt is made to find points
that lie just beyond the current exclusion limits.  Instead, the set of
benchmark points includes both points that are excluded%
\footnote{Specifically, Points 12--16, 30, 40, and 50, according to
\texttt{SModelS 2.0.0}~\cite{Kraml:2013mwa,Ambrogi:2017neo,Dutta:2018ioj,Heisig:2018kfq,Ambrogi:2018ujg,Khosa:2020zar,Alguero:2020grj,Alguero:2021dig}.}
by direct \ac{SUSY} searches in Run 2 of the
LHC and points that remain allowed, thus ensuring that the parameter
space region containing the benchmark points is relevant for Run 2 and
Run~3. In general we expect the signal to be harder to separate from the background for points that are not excluded yet;
this can be due to smaller SUSY production sections, smaller
branching ratios for decays leading to the considered signature, or
more similar kinematic features of signal and background events.
Point 0 and Point 12 are used for comparing a cut-and-count analysis and
\ac{ML} approaches.

\begin{table}
    \centering
    \caption{\label{tab:point0_mass}Masses of \ac{SUSY} and Higgs particles for
    parameter space Point 0.}
    \vspace{-2ex}
    \begin{tabular}{ccc}
        \begin{tabular}[t]{|cS|}
            \hline
            Particle & {Mass [GeV]} \\
            \hline
            ${\stauone}$  & 314.3 \\
            ${\nustau}$  & 304.2 \\
            ${\tilde{\tau}_2}$ & 2480.8 \\
            ${\se}$  & 369.5 \\
            ${\nuse}$  & 361.0 \\
            ${\tilde{e}_R}$ & 1078.2 \\
            ${\smu}$  & 369.1  \\
            ${\nusmu}$  & 360.6 \\
            ${\tilde{\mu}_R}$ & 1078.0 \\
            ${\tilde{g}}$ & 2729.5 \\
            \hline
        \end{tabular} &

        \begin{tabular}[t]{|cS|}
            \hline
            Particle & {Mass [GeV]} \\
            \hline
             ${\tilde{t}_1}$ & 7847.1 \\
             ${\tilde{t}_2}$ & 8195.9 \\
             ${\tilde{b}_1}$ & 7846.2 \\
             ${\tilde{b}_2}$ & 8051.8 \\
             ${\tilde{u}_L}$ & 2382.8 \\
             ${\tilde{u}_R}$ & 2099.4 \\
             ${\tilde{d}_L}$ & 2384.0 \\
             ${\tilde{d}_R}$ & 2310.9 \\
             ${\tilde{c}_L}$ & 2382.8 \\
             ${\tilde{c}_R}$ & 2099.4 \\
             ${\tilde{s}_L}$ & 2383.9 \\
             ${\tilde{s}_R}$ & 2310.9 \\
            \hline
        \end{tabular} &

        \begin{tabular}[t]{|cS|}
            \hline
            Particle & {Mass [GeV]} \\
            \hline
            ${\chione}$  & 360.5 \\
            ${\tilde{\chi}^0_2}$  & -375.6  \\
            ${\tilde{\chi}^0_3}$  &  527.9 \\
            ${\tilde{\chi}^0_4}$  &  995.8 \\
            ${\tilde{\chi}^{\pm}_1}$  & 369.1  \\
            ${\tilde{\chi}^{\pm}_2}$  & 995.8  \\
            $h$  & 123.9  \\
            $H$  & 4475.1  \\
            $A$  & 4475.1  \\
            $H^{\pm}$ & 4457.6   \\
            \hline
        \end{tabular}
    \end{tabular}
\end{table}

Point 0 is a benchmark for a scenario that is very difficult to detect.  Its
mass spectrum is shown in \cref{tab:point0_mass}.  The SUSY production modes
are summarized in \cref{tab:point0_xsec}.  The dominant production modes are
chargino$\,+\,$neutralino and chargino$\,+\,$char\-gi\-no with a combined cross
section of $\SI{87.5}{fb}$. The first two generations of sleptons are
produced at a much lower rate of $\SI{16.8}{fb}$ in total. Direct production
of $\stauone$ and $\nustau$ does not lead
to final states considered in the given analysis (further discussed in~\cref{sec:evtsel}).

The mass hierarchy among the lighter superpartners is
\begin{equation*}
m_{\chitwo}, m_{\se}, m_{\smu}, m_{\tilde{\chi}^{\pm}_1}, m_{\nuse}, m_{\nusmu},
m_{\chione} > m_{\stauone} > m_{\nustau} \,,
\end{equation*}
where the masses are arranged in descending order, and the
masses of $\tilde{\chi}^{\pm}_1$, $\chione$, $\chitwo$, and the first two
generations of sleptons are all within $\SI{15}{GeV}$ of each other.
This limits the available decay modes, which are listed in detail in
\cref{tab:point0_brs}.
The lighter chargino decays predominantly into $\nustau \tau$, with a
contribution of about $6\%$ from $\stauone \nu_\tau$. The
lighter neutralinos decay into $\nustau \nu_{\tau}$ and $\stauone \tau$,
with the former mode dominating for $\chione$ and the latter for~$\chitwo$.
The lighter selectrons and smuons
$\tilde{\ell}_L$ decay into $\chione\ell$ over $95\%$ of the time.%
\footnote{Here and in the following $\ell$ represents $e$ and $\mu$.}
The sneutrino $\nuse$ decays primarily into $\chione \nu_e$, with the
three-body decay $\nustau e \tau$ happening only $4\%$ of the time.
However, for $\nusmu$ the $\chione \nu_{\mu}$ decay is suppressed due to
the masses being very close to each other; this leads to the $\nustau
\mu \tau$ decay occurring with a branching ratio of roughly $50\%$.
Finally, $\stauone$ can decay into
$\nustau \ell \nu_\ell$, $\nustau \tau \nu_\tau$,
$\nustau d \bar{u}$, and $\nustau s \bar{c}$,
where the decays producing quarks dominate with a branching ratio of
around $70\%$.
\begin{table}[tbp]
    \caption{\label{tab:point0_xsec}Dominant SUSY production modes
    for Point 0.  Only channels contributing more than $1\%$ are included.}
    \begin{tabular}{lccccccccc}
        \toprule 
	    Electroweakino & $\chione \chionep $ & $\chione \chionem $ & $\chionep \chionem$ &
        $\chitwo \chionep$ & $\chitwo \chionem$ & $\chitwo \chione$ & & & Total\\
        Cross section [fb] & $ 19.4 $ & $9.3$ & $16.2  $ & $18.3$ & $8.7 $ &
        $14.5$ & & & $87.5$\\
        \midrule
	    Slepton & $\sep \nuse $ & $\smup \nusmu $ & $\sem \nusebar $ & $\smum
        \nusmubar $ & $\nuse \nusebar$ & $\nusmu \nusmubar$ & $\sem \sep $ & $
        \smum \smup$ & Total\\
        Cross section [fb] & $3.7$ & $3.7$ & $1.6$ & $1.7$ & $1.4$ & $1.4$ &
        $1.4$ & $1.4$ & $16.8$\\ 
        \bottomrule
    \end{tabular}
\end{table}

The mass spectrum of uncolored superparticles for Point 12 is relatively
close to the one considered in the analysis of ref.~\cite{Figy:2010hu},
but with parameters adjusted to obtain the right Higgs mass.  The full
spectrum is shown in \cref{tab:point12_mass}.  Point 12 serves as a
benchmark for a point with a high signal cross section; in fact, it is now
excluded by an ATLAS search for direct stau production \cite{Aad:2019byo},
as determined by recasting the results of that analysis using
\texttt{SModelS 2.0.0}~\cite{Kraml:2013mwa,Ambrogi:2017neo,Dutta:2018ioj,Heisig:2018kfq,Ambrogi:2018ujg,Khosa:2020zar,Alguero:2020grj}.
The
dominant production modes for superparticles are summarized in
\cref{tab:point12_xsec}.  The cross section for the production of
charginos and neutralinos is $\SI{457.2}{fb}$, much higher than for
Point 0.  The production of the heavier neutralino $\chitwo$ is
irrelevant here because it is much heavier than $\chione$.
The production of first- and second-generation sleptons has a cross
section of $\SI{100.2}{fb}$. Again, direct $\stauone$ and $\nustau$
production rates are negligible.

\begin{table}[btp]
    \centering
    \caption{\label{tab:point12_mass}Masses of \ac{SUSY} and Higgs particles for Point 12.}
    \vspace{-2ex}
    \begin{tabular}{ccc}
        \begin{tabular}[t]{|cS|}
            \hline
            Particle & {Mass [GeV]} \\
            \hline
             ${\stauone}$  & 119.1  \\
             ${\nustau}$  & 90.7 \\
             ${\tilde{\tau}_2}$ & 193.2 \\
             ${\se}$  &  225.1 \\
             ${\nuse}$  & 211.4 \\
             ${\tilde{e}_R}$ & 1202.0 \\
             ${\smu}$  & 249.1 \\
             ${\nusmu}$  & 236.7 \\
             ${\tilde{\mu}_R}$ & 1391.7 \\
             ${\tilde{g}}$ & 2552.7 \\
            \hline
        \end{tabular} &

        \begin{tabular}[t]{|cS|}
            \hline
            Particle & {Mass [GeV]} \\
            \hline
             ${\tilde{t}_1}$ & 2174.0 \\
             ${\tilde{t}_2}$ & 4885.4 \\
             ${\tilde{b}_1}$ & 3537.1 \\
             ${\tilde{b}_2}$ & 4880.2 \\
             ${\tilde{u}_L}$ & 2429.7 \\
             ${\tilde{u}_R}$ & 1658.6 \\
             ${\tilde{d}_L}$ & 2430.8 \\
             ${\tilde{d}_R}$ & 1771.8 \\
             ${\tilde{c}_L}$ & 3008.2 \\
             ${\tilde{c}_R}$ & 2066.3 \\
             ${\tilde{s}_L}$ & 3009.0 \\
             ${\tilde{s}_R}$ & 2234.5 \\
            \hline
        \end{tabular} &

        \begin{tabular}[t]{|cS|}
            \hline
            Particle & {Mass [GeV]} \\
            \hline
            ${\chione}$  & 278.4 \\
            ${\tilde{\chi}^0_2}$  & 726.7 \\
            ${\tilde{\chi}^0_3}$  & -765.0 \\
            ${\tilde{\chi}^0_4}$  & 797.6 \\
            ${\tilde{\chi}^{\pm}_1}$  &  278.7 \\
            ${\tilde{\chi}^{\pm}_2}$  & 771.6 \\
            $h$  & 125.5  \\
            $H$  & 4741.3  \\
            $A$  & 4741.2  \\
            $H^{\pm}$ & 4742.6   \\
            \hline
        \end{tabular}
    \end{tabular}
\end{table}

At Point 12, the mass hierarchy among the lighter sparticles is
\begin{equation*} m_{\tilde{\chi}^{\pm}_1} , m_{\chione} >  m_{\smu} >
m_{\nusmu} > m_{\se} > m_{\nuse} >  m_{\stauone} > m_{\nustau} \,.
\end{equation*} Unlike Point 0, here $m_{\chione} > m_{\tilde\ell_L},
m_{\tilde\nu_{\ell}}$, allowing for a wider range of decay modes and forcing
the first two generations of sleptons into three-body decay modes.  The
branching ratios are summarized in \cref{tab:point12_brs}.  The lightest
neutralino has the decay modes $\stauone \tau$, $\nustau \nu_\tau$,
$\tilde\ell_L \ell$, and $\tilde\nu_\ell \nu_\ell$.  The decays are dominated
by the $\stauone$ and $\nustau$ channels with branching ratios of about $35\%$
and $40\%$, respectively.  The decay modes of the lighter chargino
$\tilde\chi^\pm_1$ are $\stauone \nu_{\tau}$, $\nustau \tau$, $\tilde\nu_\ell
\ell$, and $\tilde\mu_L \nu_\mu$, with the $\stauone$ and $\nustau$ decays
contributing $35\%$ and $45\%$, respectively.  Decays into $\tilde\nu_\ell$
account for another $15\%$ while $\tilde\ell_L$ modes are heavily suppressed.
The first two generations of sleptons $\tilde\ell_L$ have the decay modes
$\nustau \tau \nu_\ell$, $\stauone \tau \ell$, and $\nustau \ell \nu_\tau$,
where a tau is produced in almost $80\%$ of the decays.  Electron and muon
sneutrinos decay into $\nustau \tau \ell$, $\nustau \nu_\tau \nu_\ell$,
$\stauone \ell \nu_\tau$, and $\stauone \tau \nu_{\ell}$, where the branching
ratios are about $10\%$ for $\stauone \tau \nu_{\ell}$ and about $30\%$ for
each of the other three decay modes.  The $\stauone$ decay modes are virtually
the same as for Point~0.

\begin{table}[tb]
    \caption{\label{tab:point12_xsec}Dominant SUSY production modes
    for Point 12. Only channels contributing more than $1\%$ are included.}
    \begin{tabular}{lccccccccc}
        \toprule
        Electroweakino & $\chione \chionep $ & $\chionem \chione$ & $\chionep
        \chionem$ & &&&&& Total\\
        Cross section [fb] & $199.2$ & $108.5$ & $147.1 $ & &&&&& $457.2$ \\
        \midrule
        Slepton & $\sep \nuse $ & $\smup \nusmu $ & $\sem \nusebar $
        & $\smum \nusmubar $ & $\nuse \nusebar$ & $\nusmu \nusmubar$ & $\sem
        \sep $ & $ \smum \smup$ & Total\\ 
        Cross section [fb] & $23.7$ & $14.9 $ & $13.3$ & $7.31$ & $11.9$ & $6.8$ &
        $8.7$ & $5.9$ & $100.2$\\
        \bottomrule
    \end{tabular}
\end{table}

For both Point 0 and Point 12, taus are quite likely to be produced in the
prompt sparticle decay chains, since these end in the $\nustau$ NLSP, in
particular from neutralinos and charginos, whose production cross sections
dominate.  In order to arrive at the signature of two taus and one muon, which
we will use in the following, an additional muon is required.  This muon can be
produced in slepton decays.  However, this happens only in about $10\%$ of
$\stauone$ decays.  Decays of first- and second-generation sleptons are more
likely to yield muons, but these sleptons are unlikely to arise from neutralino
and chargino decays and have a relatively small direct production cross
section. Depending on the point considered it can suppress or enhance the overall yields greatly.

Comparing the two parameter space points, an important difference is obviously
the larger SUSY production cross section for Point 12.  However, for our 
analysis it turns out to be more important that the Point 0 mass spectrum of
the lighter superparticles is much more compressed and has a neutralino
$\chione$ that is lighter than the first- and second-generation sleptons, which
leads to very different decay chains.  In particular, the lighter charginos and
neutralinos can decay to first- and second-generation sleptons and leptons with
a branching ratio of more than $1\%$ only for Point 12.  In combination, these
factors lead to a much higher yield of events with two taus and one muon for
Point 12 than for Point 0, see~\cref{tab:parampoints_yields} below.

\FloatBarrier 

\subsection{Event generation}
\label{sec:evtgen}
The events are assumed to be produced in proton-proton collisions at a
center-of-mass energy of \SI{13}{\TeV}. Monte Carlo \ac{SM} background events
are generated by \texttt{SHERPA 2.2.4}~\cite{Bothmann:2019yzt}, using
the \texttt{NNPDF3.0} \cite{Ball_2015} parton distribution functions (PDF) set
and $\alpha_s(M_Z)=0.118$.
The types of background processes and corresponding numbers of events are
given in \cref{tab:bkg} below.

The \ac{SUSY}
signal events are produced with \texttt{Herwig 7.1.3}~\cite{Bahr:2008pv,Bellm:2015jjp} at the leading order.
A computation by \texttt{Prospino2}~\cite{Beenakker:1999xh} shows that the
next-to-leading-order cross section is about $25\%$ larger.  Importantly
for our analysis, however, the kinematical distributions of uncolored
final-state particles are not drastically altered by higher-order
effects~\cite{Baglio:2016rjx,Fiaschi:2019zgh}, especially for the
dominant electroweakino production modes.
For simplicity, only production via gluon-gluon fusion and quark annihilation is
considered, which are the dominating production modes at a proton collider.

For the signal generation the
\texttt{MMHT2014}~\cite{Harland_Lang_2015} PDF set is used, as it is the
default for version 7.1 of \texttt{Herwig}. Also here the strong
coupling $\alpha_s(M_Z)$ is set equal to $0.118$. A comparison with
alternative PDF sets, \texttt{CT14} \cite{Dulat_2016} and
\texttt{NNPDF3.0}, is performed for signal samples and used as an
uncertainty. The \texttt{MMHT2014} PDF set results in the most
conservative cross section prediction (with a difference of up to $10\%$),
while the modelling of the kinematic variables remains consistent. The
values of renormalization and factorization scales are varied by a factor of $2$ and the difference with the nominal is used as an additional systematic uncertainty. The effect on the shape of the kinematic variables and overall normalization has been found to be negligible.

All relevant two- and three-body sparticle decays are included.  Their
branching ratios are computed by \texttt{SPheno 4.0.3}.

\subsection{Event selection} \label{sec:evtsel}
For the purpose of the analysis it is assumed that the events are recorded by a
general-purpose particle detector like CMS~\cite{Chatrchyan:2008} or
ATLAS~\cite{Aad:2008}.  The definitions of physical objects reflect the usual
selection criteria used by such detectors.  In particular, this implies an
upper limit on the pseudorapidity $|\eta|$ to match the typical detector
geometry. \texttt{Rivet 2.5.4}~\cite{Buckley:2010ar} is used for event
selection and object definitions.  

Jets are reconstructed using the anti-$k_T$ clustering algorithm
\cite{Cacciari:2008gp} with distance parameter $R=0.4$ implemented via the
\texttt{FastJet} package~\cite{Cacciari:2011ma, Cacciari:2005hq}. They are
required to have transverse momentum $\pt \geq \SI{20}{GeV}$ and $|\eta| <
2.8$.  Electrons, muons and taus are required to have \pt\ of at least
$\SI{15}{GeV}$ and $|\eta| < 2.5$.  Note that we treat only hadronically
decaying taus as physical objects.  For leptonic decays, the daughter particles
are considered instead.  The reason for this separation is that it is generally
hard to identify leptonically decaying tau leptons as such in proton-proton
collisions. 

An overlap removal procedure is applied to all events to mirror what would be
done when dealing with real data.  When multiple objects are reconstructed from
the same detector signature all but one are ignored.  This is done to improve
the likeness of simulated events to what could be seen in an experiment and to
make sure that the training of the \ac{ML} models excludes features that are
only accessible in Monte Carlo events.  Even if the overlapping objects are
real this information is not available to a detector and hence all but one of
them are removed.  The successive steps of the overlap removal procedure are
summarized in~\cref{tab:OR}.

\begin{table}[tb]
  \centering
 \caption{Steps of the overlap removal algorithm.
    If two or more different objects are separated by less than the matching condition,
    only one is retained, according to these rules. For example, if a muon and a jet 
    are separated by $\Delta R = 0.1$, rule 2 is invoked and the muon is kept; if the
    separation is $0.3$, rule 5 is invoked and the jet is kept. Only surviving objects
    participate in subsequent steps. }
    \begin{tabular}{clll}
         & Object discarded          & Object kept      & Matching condition      \\
    \midrule
    1.   & jet                       &  electron        & $\Delta R < 0.2$        \\
    2.   & jet                       &  muon            & $\Delta R < 0.2$        \\
    3.   & jet                       &  had.~tau        & $\Delta R < 0.2$        \\
    4.   & electron                  &  jet             & $\Delta R < 0.4$        \\
    5.   & muon                      &  jet             & $\Delta R < 0.4$        \\
    6.   & had.~tau                  &  jet             & $\Delta R < 0.4$        \\
    7.   & electron                  &  had.~tau        & $\Delta R < 0.4$        \\
    8.   & muon                      &  had.~tau        & $\Delta R < 0.4$        \\
  \bottomrule
  \end{tabular}
   \label{tab:OR}
\end{table}

Events that contain at least two hadronically decaying tau leptons with the same
electric charge and a muon of opposite charge are selected. 
The signature $\tau_h^{\pm} \tau_h^{\pm} \mu^{\mp}$ is used for several
reasons. First of all,
$\nustau$ being the \ac{NLSP} leads to a plethora of decay modes of \ac{SUSY}
particles with tau leptons in the final states, see above and
\cref{tab:point0_brs,tab:point12_brs}.
Secondly, a three-lepton signature with same-sign same-flavour leptons
heavily reduces \ac{SM} background. This is especially important to
suppress \ac{SM} events with $Z$ boson production in association with jets 
while not particularly hurting the signal yields. The
$\tau_h^{\pm} \tau_h^{\pm} \mu^{\mp}$ signature is chosen over, e.g., $\mu^{\pm}\mu^{\pm}\tau_{h}^{\pm}$ because the signal-to-background ratio is higher.
The three lepton requirement by itself is also very effective at 
suppressing the production of a $W$ boson in association with jets.
Requiring only one or two leptons would lead to an explosive growth of the \ac{SM} 
background. The study presented is inclusive to events with four or more
leptons, but due to the low yields of such processes it would not be
beneficial to require more than three leptons. 

Finally, in the context of this study there is no particular difference
in whether an electron or a muon is used in the final state. The
\ac{SUSY} yields might change slightly depending on the parameter space
point, but the methodology (which is our main focus) remains the same.
Muons are chosen as the default as they are generally easier to detect. 

For each parameter space point, \cref{tab:parampoints_yields}  shows the
expected number of signal events satisfying all requirements described
in this section (i.e., after overlap removal the events contain
$\tau_h^\pm \tau_h^\pm \mu^\mp$ with $\pt > \SI{15}{GeV}$ and
$|\eta|<2.5$).  In addition to the total expected yield, the table
contains the number of events from each production mechanism for the
original superparticles.

\begin{table}[tb]
    \centering 
    \caption{Expected number of signal events for the parameter points used in
    the analysis after the event selection described in \cref{sec:evtsel}
    is applied.  An integrated luminosity of $\SI{149}{fb^{-1}}$ is assumed.
    The last three columns specify the production type of the SUSY particles,
    viz.\ slepton, electroweakino (neutralino and chargino), and strong (squark
    and gluino) production.
    }
    \begin{tabular}{lSSSS}
        & {Slepton} & {Electroweakino} & {Strong} & {Total yield} \\
        \midrule
        Point 0  & 14.9  & 7.6   & 1.4 & 23.9   \\
        Point 12 & 248.1 & 270.1 & 3.6 & 521.8  \\
        Point 13 & 282.5 & 350.9 & 2.7 & 636.1  \\
        Point 14 & 305.5 & 385.7 & 0.5 & 691.7  \\
        Point 15 & 442.9 & 612.1 & 1.7 & 1056.7 \\
        Point 16 & 421.5 & 878.4 & 6.9 & 1306.8 \\
        Point 20 & 13.2  & 30.2  & 1.2 & 43.6   \\
        Point 30 & 19.5  & 26.1  & 0.3 & 45.9   \\
        Point 40 & 439.2 & 471.2 & 4.4 & 914.8  \\
        Point 50 & 841.5 & 726.4 & 1.3 & 1569.2 \\
        \bottomrule
    \end{tabular}
	\label{tab:parampoints_yields}
\end{table}

\section{Cut-based analysis}
\label{sec:cut_analysis}
Simple cut-and-count analyses are performed on parameter space Point 0 and
Point 12 to serve as a baseline for the evaluation of various \ac{ML} methods.
The selections used for the analyses are optimized by maximizing the
statistical significance $z$ defined as
\begin{equation}
    \label{eq:significance}
    z = \sqrt{2 \Bigl[(S+B) \, \ln\frac{S+B}{B} - S \Bigr] } \,,
\end{equation}
where $B$ is the theoretical prediction for the number of \ac{SM} background
events and $S+B$ is the observed yield, or sum of the theoretical signal and
\ac{SM} backgrounds. The kinematic variables are scanned in significance, i.e., a cut maximizing $z$ is selected for a given 
variable. This procedure is performed sequentially for all variables considered. 
No further correlation information is used.
This is done to contrast with the \ac{ML} models that typically do take the 
correlation between input variables into account.

We are assuming $\SI{149}{fb^{-1}}$ integrated luminosity to determine the
numerical values of $S$ and $B$. This choice is motivated by the integrated
luminosity recorded by the ATLAS and CMS experiments during the 2015--18
proton-proton collisions at $\sqrt{s} = \SI{13}{TeV}$ (Run 2). The expected
background yields before any optimization is applied are summarized
in~\cref{tab:bkg}. Note that $W/Z+$jets production is heavily suppressed by the event selection and is expected to contribute less than $0.1\%$ of the total background
yields. Therefore, it is not considered further on. Similar comments apply to multijet production.

We do not expect significant contributions from fake taus given that
sufficiently tight tau reconstruction and identification algorithms are
used in actual analyses.  A precise fake tau background estimation would
have to be data-driven and experiment-specific, as Monte Carlo
generators are not always reliable for modeling fake taus.

\begin{table}[tb]
    \centering
    \caption{Expected number of background events after the event selection
    described in \cref{sec:evtsel} is applied. An integrated luminosity
    of $\SI{149}{fb^{-1}}$ is assumed.} 
    \begin{tabular}{lS}
        {Events} & {Expected yield} \\
        \midrule
        ttZ  & 133 \\
        ttW  & 82 \\
        ZZ  & 189 \\
        WZ  & 443 \\
        WWW  & 11 \\
        ZWW  & 11 \\
        WWbb  & 6094 \\
    \midrule
        Total & 6963 \\
    \bottomrule
    \end{tabular}
    \label{tab:bkg}
\end{table}

\subsection{Input features} \label{sec:input_feat}
The same discriminating variables are used for the cut-and-count analysis and
for the training of the \ac{ML} methods. The selection is based on preliminary
studies to optimize the number of necessary input features. These variables
include \pt, $\eta$ and azimuthal angle $\phi$ of the three objects used for
the event selection -- the two hadronically decaying tau leptons with the same
charge and the muon with the opposite charge. The physical objects are ranked
by \pt; whenever ``leading'' or ``second'' tau lepton is mentioned, it is in
this context. If more than one muon satisfying the selection criteria is
present only the one with the highest \pt\ is used.  The absolute value \met\
and the azimuthal angle \metphi\ of the missing transverse momentum are also
included in the input features.  Finally, the scalar sum of the transverse
momenta of all visible objects in the event, \HT, and the numbers of jets,
hadronically decaying tau leptons, electrons and muons in the event are used.
The list of variables used is summarized in~\cref{tab:input_feat}.

The cut-and-count analysis relies on constructing combinatorial
variables that are commonly used in high energy physics searches, such
as the angular difference between two objects.  These ``advanced''
variables are not used as input for the \ac{ML} training as it is assumed
that a sufficiently sophisticated algorithm should be able to achieve
the same (or better) performance based on the basic input variables alone.

\begin{table}[tb]
    \centering
    \caption{ \label{tab:input_feat} Overview of the input features used for
    the cut-based analysis and for the training of the \ac{ML} models.}
    \begin{tabular}{cl}
        Object & Variables\\
        \midrule
        Leading tau & \pt, $\phi$, $\eta$ \\
        Second tau  & \pt, $\phi$, $\eta$ \\
        Muon        & \pt, $\phi$, $\eta$ \\
        Missing momentum   & \met, \metphi \\
        Transverse momenta & \HT \\
        Leptons and jets   & $n_e$, $n_\mu$, $n_\tau$, $n_\text{jet}$\\
        \bottomrule
    \end{tabular}
\end{table}

\subsection{Point 0}
\label{sec:point0}
Point 0 is in a ``hard'' part of the parameter space with only around 25 signal
events expected on top of 6963 background events after the initial event selection (as described in \cref{sec:evtsel}). A scan
in significance is performed for all input features described in
\cref{tab:input_feat}.  The variables $\phi$ and $\eta$ are not particularly
interesting by themselves, so a scan over the angles between physical objects
$\Delta \phi$ and $\Delta R$ is performed instead. In addition, we scan over
the transverse masses of hadronically decaying tau leptons defined as
\begin{equation} \label{eq:mT}
    m_T^{\tau} =
    \sqrt{2 \pt^\tau \met \left[ 1 - \cos(\phi^\tau-\metphi) \right]}
    \,.
\end{equation}

\begin{figure}[htb]
  \centering
  \begin{subfigure}{.49\linewidth}{\includegraphics[width=1.0\textwidth]{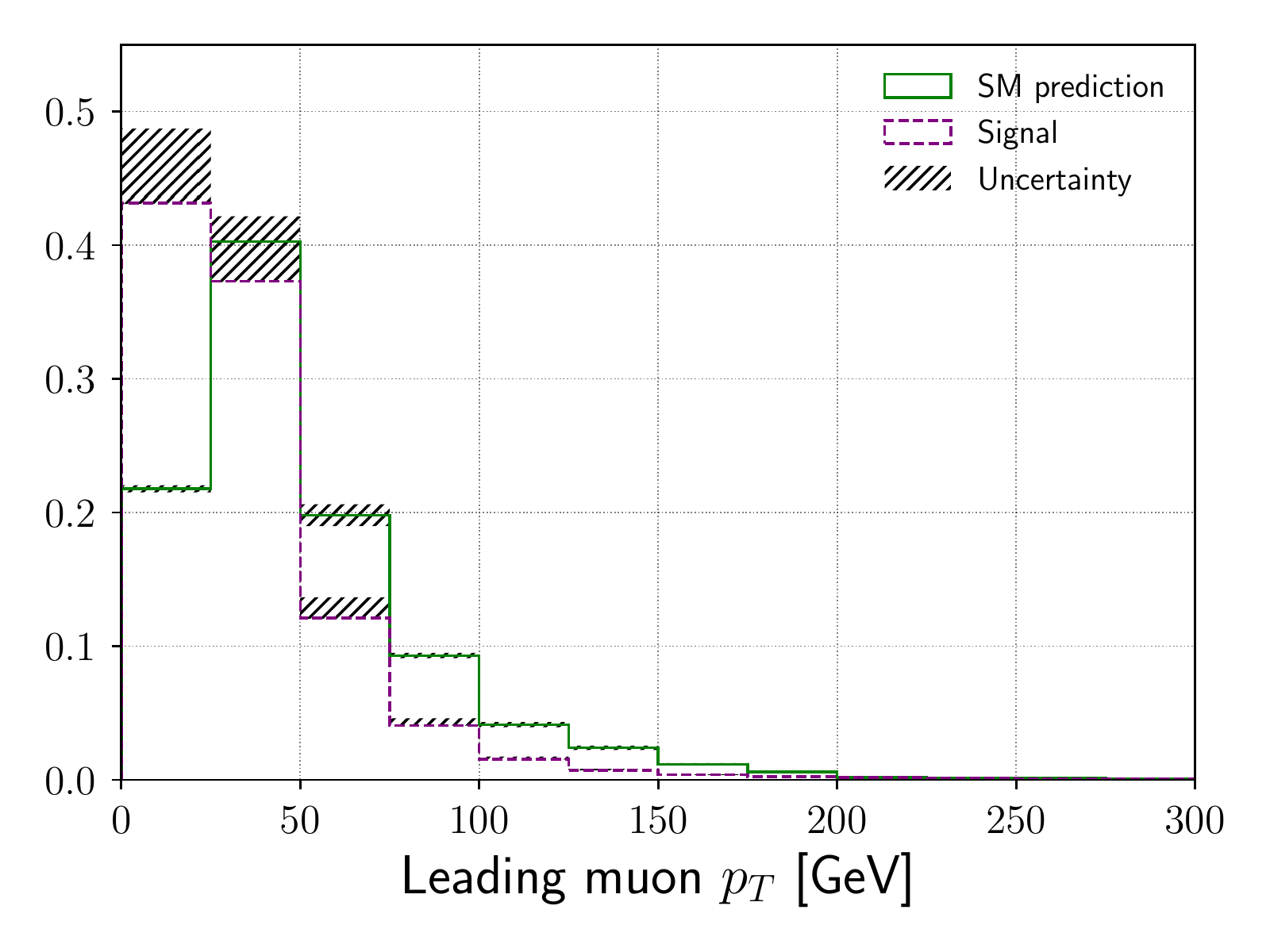}}
  \end{subfigure}
  \begin{subfigure}{.49\linewidth}{\includegraphics[width=1.0\textwidth]{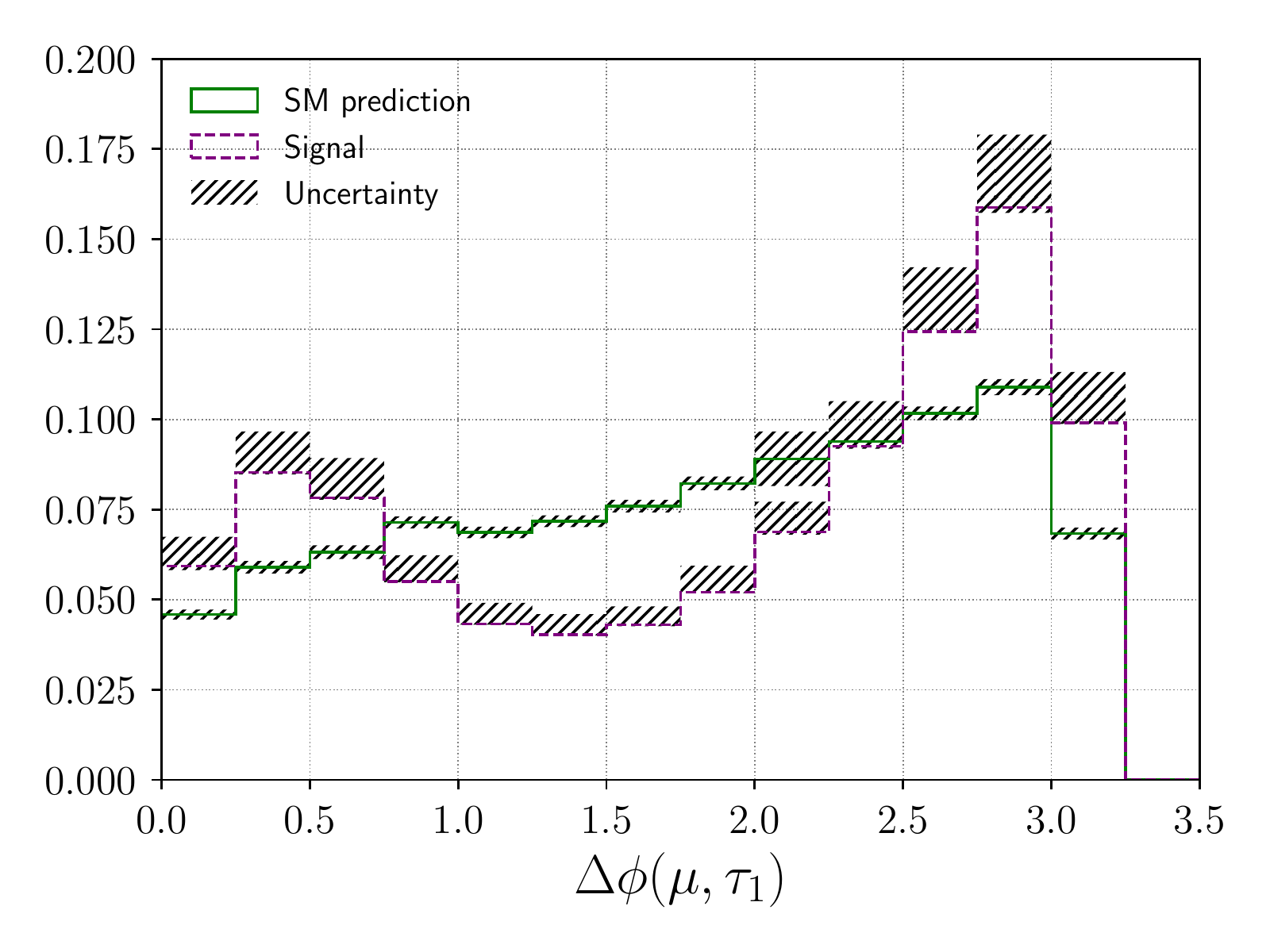}}
  \end{subfigure}
    \caption{\label{fig:point0_cut}Distributions of various variables for \ac{SM}
    background and Point 0 \ac{SUSY} signal, normalized to $1$.
    The hatched bands show the combined statistical and theoretical uncertainties of the signal and the statistical uncertainty of the background. 
    The asymmetry of the signal uncertainty stems from the comparison with alternative PDF sets.
	}
\end{figure}

The simple cut-based approach is not appropriate for Point 0 due to the
extremely low number of expected signal events and the difficulty in
reducing the number of background events. While there are noticeable
differences between the signal and the backgrounds in the distributions
for some of the variables, see~\cref{fig:point0_cut} for two examples,
there are no obvious selection criteria that could efficiently exploit
these differences.  As a result, no tightening in selections leads to an
increase over the nominal significance of $z=0.3$.  However, an
improvement is expected with the \ac{ML} methods as they should be better at
exploiting the signal-background differences.

While the significance of $z=0.3$ might seem completely hopeless at the first
glance, it should be noted that $z$ scales with the square root of the
luminosity. Combining Run 2 with the future Run 3 would double the expected
integrated luminosity. The high-luminosity LHC project \cite{Aberle:2749422} aims to increase the
current LHC luminosity by a factor of 10 and to bring the total integrated luminosity yields up to
$\SI{4000}{fb^{-1}}$ of data. Although the conditions during Run~3 will be
different from Run 2, this is, by the numbers, already almost enough for exclusion by
itself. If the \ac{ML} methods can improve the sensitivity even slightly, Point
0 is worth considering.

Note also that the \emph{absolute} significance determined by our
analyses is subject to considerable uncertainties (e.g., due to the lack
of a detector simulation and the generation of signal events at leading
order), since our focus is on comparing different methods and thus on
\emph{relative} values, where uncertainties are expected to cancel out
to a large degree.

\subsection{Point 12}
\label{sec:point12}
Point 12 is comparatively ``easy'' to detect. More than 520 signal events are
expected for this point on top of 6963 background events. This is already
enough to reach $5 \sigma$ discovery significance by itself.  Hence, it is not
surprising that ref.~\cite{Aad:2019byo} was able to rule out this point.

\begin{figure}[htb]
  \centering
  \begin{subfigure}{.49\linewidth}{\includegraphics[width=1.0\textwidth]{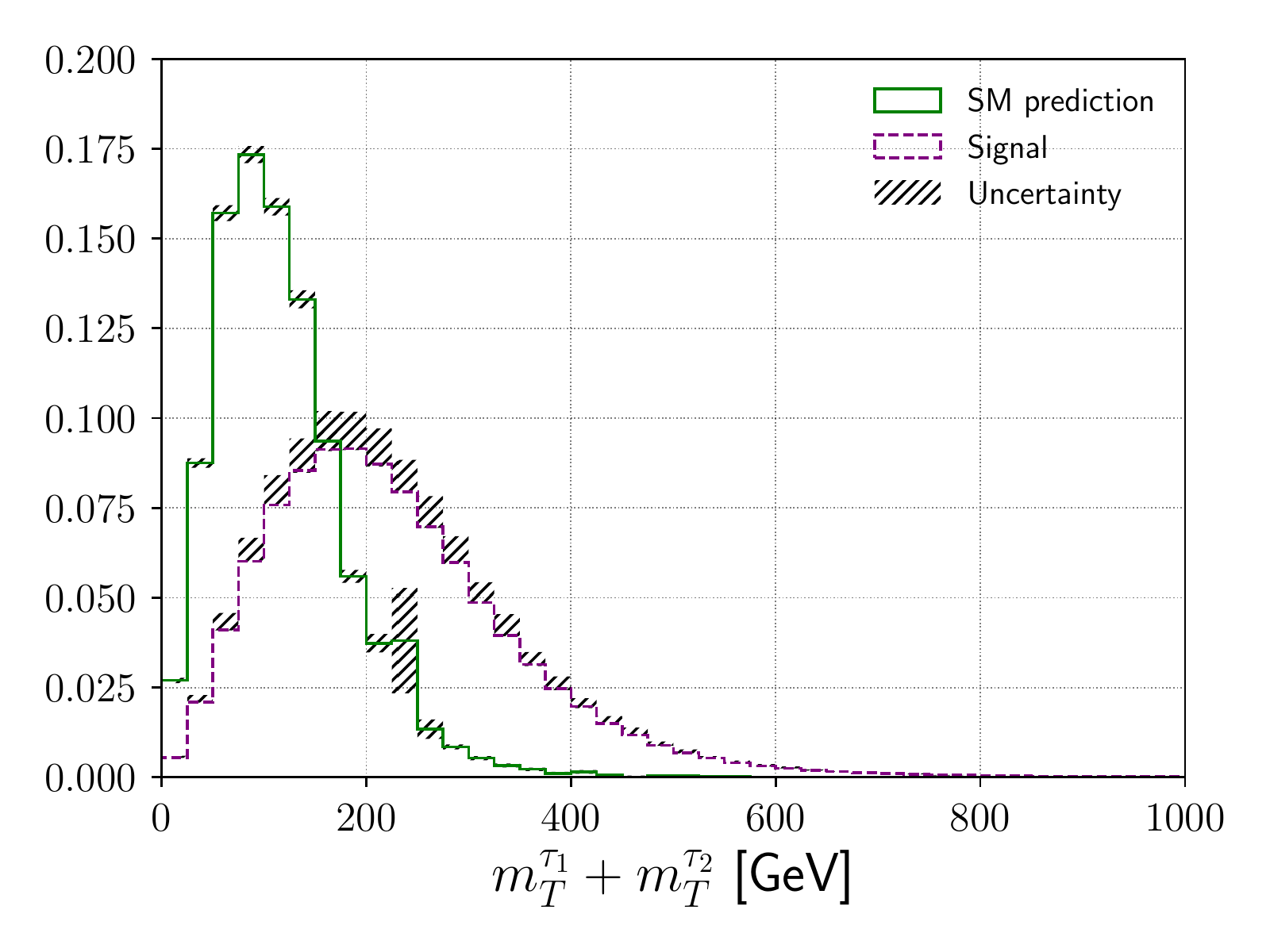}}
  \end{subfigure}
  \begin{subfigure}{.49\linewidth}{\includegraphics[width=1.0\textwidth]{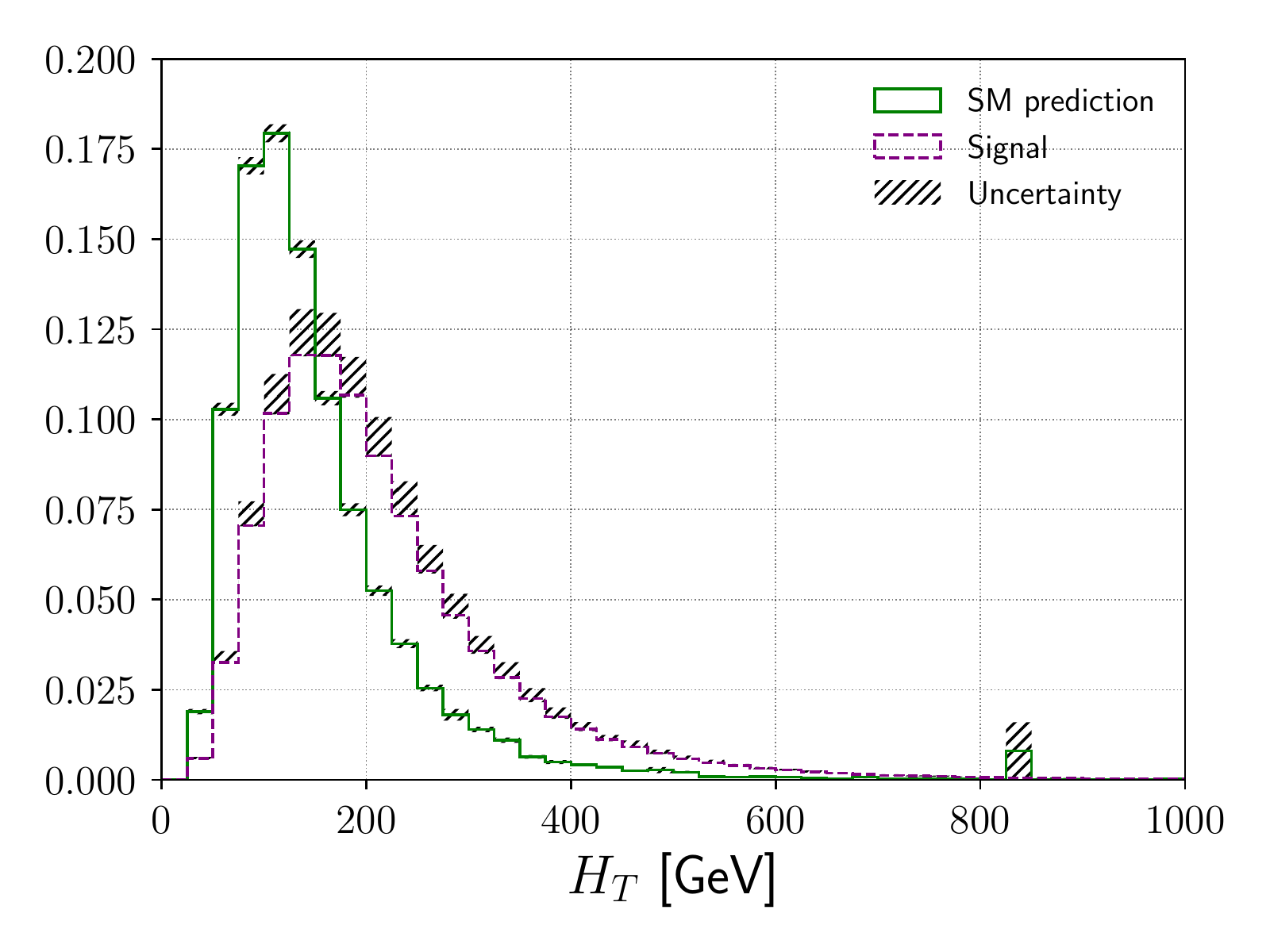}}
  \end{subfigure}
    \caption{\label{fig:point12_cut}Distributions of variables used for the
    cut-based analysis, \ac{SM} background and Point 12 \ac{SUSY} signal,
    normalized to 1. The hatched bands show the combined statistical and theoretical uncertainties of the signal and the statistical uncertainty of the background.
    The asymmetry of the signal uncertainty stems from the comparison with alternative PDF sets.}
\end{figure}

Significance scans are performed on all the input features, on the tau
lepton transverse masses $m_T^\tau$, and on the angular variables
$\Delta\phi$ and $\Delta R$.  The selection maximizing the significance
includes requiring $H_T > \SI{125}{GeV}$ and the sum of the tau lepton
transverse masses to be larger than $\SI{250}{GeV}$, see
\cref{tab:point12_cut}.  Plots of signal and background variables used for the
selection (both normalized to 1) are presented in~\cref{fig:point12_cut}.
After the optimization we expect 143 background and 148 signal events,
corresponding to $z_{\text{Cut\&count}} = 10.8$. 

\begin{table}[tb]
    \centering
    \caption{Final selection for the cut-based analysis of parameter space Point 12.}
    \begin{tabular}{cl}
        Variable & Cut\\
        \midrule
        \HT  & $\geq \SI{125}{GeV}$ \\
        $m_T^{\tau_1} + m_T^{\tau_2}$ & $\geq \SI{250}{GeV}$ \\
        \bottomrule
    \end{tabular}
    \label{tab:point12_cut}
\end{table}

\subsection{Tau selection efficiency} \label{sec:TauID}
Initial event selection requires at least two hadronically decaying tau
leptons. It is important to notice that at hadron colliders the tau selection
efficiency $\epsilon_\tau$ can be significantly lower than $1$ depending on the
desired purity and rejection rates \cite{ATL-PHYS-PUB-2019-033}. It follows
that the significance values quoted should be scaled by a factor of
$\sqrt{\epsilon_\tau \cdot \epsilon_\tau} = \epsilon_\tau$ to obtain a
realistic estimate of what can be achieved at a hadron collider. This is not
particularly important for the comparison of different algorithms and we assume
$\epsilon_\tau = 1$ everywhere for simplicity. However when deciding whether a
point in the parameter space can be tested in an experiment this is a crucial
point to consider.

\section{Machine learning methods}
Let $X$ be a set of observable variables in an event, such as the features
listed in~\cref{tab:input_feat}, and let $y$ be the corresponding value
representing the class of the process responsible for the event, i.e., \ac{SM}
(labeled $0$) or \ac{SUSY} (labeled $1$). In principle, for a properly selected
set $X$ there exists a mapping of $X$ to $y$. Due to the limitations of
real-life detectors, including the inability to measure stable neutral
particles, $X$ cannot be mapped to $y$ unambiguously in realistic experiments.
The relationship becomes $f(X) = \hat{y}$, where $0 \leq \hat{y} \leq 1$ is
called the \textit{predicted} value and represents the degree of certainty in
the class of the process responsible for the event.  A statistical model can
then be constructed to approximate the relationship by the function
$\hat{f}(X)=\hat{y}$.

For this model construction an \ac{ML} algorithm can be used.  There are a
variety of \ac{ML} algorithms available, suitable for different tasks, all
having parameters and hyperparameters which must be tuned to fit the task at
hand. This is the \textit{learning} part, and relevant for the present
discussion is so-called supervised learning. The learning happens during a
\emph{training} phase, for which the algorithm's parameters are usually randomly
initialized and subsequently fitted. Fitting is done using a training data
set containing input features $X$ as well as data labels $y$.  It consists in
adjusting the parameters to minimize some \emph{loss function}, whose value is
the lower the closer the output $\hat y$ is to the true label $y$.
Hyperparameters can be set manually before training, or adjusted as part of
the training process, using a hyperparameter optimization procedure, e.g.,
cross validation. The result is a parameterized \ac{ML} model, or
\emph{classifier},
and its output is referred to as a prediction.  In the following, two \ac{ML}
algorithms are considered, described in~\cref{sec:xgboost,sec:nn} after a brief
discussion of the data sets.

\subsection{Data preparation} \label{sec:datasets}
A key ingredient in \ac{ML} is the data used for training, in our case
consisting of \ac{SM} background and \ac{SUSY} signal processes.  In the
part of the parameter space relevant for our analysis, the background
processes dominate the signal processes by at least $10:1$.
Nevertheless, the same number of signal and background events
is used in the training data set, since the \ac{ML} model should manage
to identify both classes equally well. 

While the relative distribution among the processes in the background
data is known from \ac{SM} predictions, this is not the case for the
signal data, since the relative occurrence of the \ac{SUSY} processes
depends on the sparticle spectrum.  The three types of signal processes
are slepton, electroweakino and strong production.  However, our analysis
concerns a part of the parameter space where very few signal events from
strong production are expected, see~\cref{tab:parampoints_yields}.
For instance, for Point 12, only one event from strong processes is
expected per $132$ signal events. Therefore, we choose not to include events from strong production in the training data.

Furthermore, it is certainly possible to train a classifier to identify each of the three types of signal processes separately. We choose not to do so and to treat all three as a single ``signal'' class. The focus of the presented analysis is on discovery, not on classifying different channels. A further argument against doing multi-class classification is that this would add additional degrees of freedom to the final statistical analysis (discussed in~\cref{sec:mle}), and thus potentially lead to an artificially increased discovery significance.

When training a classifier to make predictions for a particular parameter point, a distribution of the signal events according to their expected yields is necessary. However, when training a classifier to be sensitive to a wide selection of parameter points, one should be as general as possible, and in this analysis, most points have yields for slepton and electroweakino production that are approximately the same (after initial event selection), see~\cref{tab:parampoints_yields}. The training data for Point 12 is therefore equally distributed between slepton and electroweakino production.

Two sets of data are generated, representing Point 12 and Point 0, respectively.
The first contains $825\,294$ background events, distributed according
to~\cref{tab:bkg}, and $825\,280$ signal events, simulated using the Point 12
parameters but equally distributed, similar to the expected yields in table 6,
among slepton and electroweakino production channels. The second training
data set contains $1\,003\,686$ background
events, distributed according to~\cref{tab:bkg}, and $1\,003\,590$ signal
events, simulated using the Point 0 parameters and distributed according to table 6. Each
of these data sets is split into two to yield one training and one
validation data set per point. The validation data set serves several
purposes.  First, it is used during the training phase of the
classifiers.  Second, it is sampled from to find the optimal classifier
output threshold in~\cref{sec:cut_classifier}.  Finally, it is used to
construct templates in~\cref{sec:mle}.

The test data sets, on the other hand, are constructed differently. One test
data set is constructed for each of our ten parameter space points, according
to the signal and background admixture predicted by the theory,
see~\cref{tab:parampoints_yields} as would be observed at colliders. Note that
the background class is the same in all parameter space points, and only the
number of signal events as well as the distribution within the signal class
vary.  Also note that the test data sets do include processes from strong
production.

\subsection{Classifiers}
We employ two different \ac{ML} algorithms that are commonly used for
classification -- XGBoost and a \ac{DNN}. The input variables used for both
are listed in \cref{tab:input_feat}.
\subsubsection{XGBoost}
\label{sec:xgboost}
XGBoost~\cite{Chen:2016:XST:2939672.2939785}
is a commonly used tree ensemble
\ac{ML} algorithm, and has become popular for being both fast and easy to use
out of the box.  We train an XGBoost\footnote{\texttt{XGBoost 1.3.3} is used
for the XGBoost implementation.} model to separate \ac{SUSY} signal events
from \ac{SM} background events, tuning its architecture and parameters using a
cross-validation search~\cite{hastie01statisticallearning}. We use a maximum 
depth of 10 and an early stopping criterion that stops the training if the loss 
does not improve for 50 rounds. This yields a model with
a \ac{ROC} \ac{AUC} \footnote{A \ac{ROC} \ac{AUC} value of $1$ is the
theoretical maximum and corresponds to a perfect classifier, while 0.5
represents random guessing.} of $0.87$ on training and test data from
Point 12, and $0.77$ for Point~0.

\subsubsection{Deep neural network}
\label{sec:nn}
A \ac{DNN} is trained to separate \ac{SUSY} signal events from \ac{SM}
background events.\footnote{\texttt{TensorFlow 2.4.1} with the \texttt{Keras}
module is used for the \ac{DNN} implementation.} The hyperparameters are again
optimized using a cross-validation search. These are the architecture-related
ones (numbers of layers and nodes per layer), batch size, dropout rate, and
learning rate.  The final architecture used has five hidden layers containing
$500, 500, 250, 100$, and $50$ nodes, respectively, a batch-size of $50$, a
dropout rate of $0.21$, and an initial learning rate of $10^{-3}$ in the Adam
optimizer algorithm. If the accuracy does not improve over 10 consecutive epochs,
the learning rate is reduced by a factor of $100$ until it reaches $10^{-7}$. 
The training process stops if there is no improvement in the validation loss
over 15 consecutive epochs. The LeakyRelu
activation function is used in the hidden layers, and the sigmoid
activation function is used in the output layer's single node, to yield
output values in the range $[0,1]$. The loss function is binary crossentropy.

The resulting model achieves a \ac{ROC} \ac{AUC} of $0.88$ on training and test
data from Point~12, i.e., approximately the same as for the XGBoost model, and
$0.83$ for Point 0.
This does not necessarily mean that the classifiers behave in the same way, merely
that the area under the curve spanned by the True Positive Rate (TPR) and False
Positive Rate (FPR) when varying the classification threshold between the
background and signal classes is the same. In fact, the results presented
in~\cref{fig:templates} show that the two classifiers' respective outputs are
not distributed equally.

\subsection{Cutting on classifier output}
\label{sec:cut_classifier}
As a first comparison to the cut analysis described in~\cref{sec:cut_analysis},
the validation data is used to determine the classifier output value which best
separates the signal from the background class, i.e., for which the discovery
significance in~\cref{eq:significance} is maximized. We refer to this value as
the optimized cutoff value. The validation data set is resampled to contain the
number of events listed in~\cref{tab:parampoints_yields}.  The significance is
calculated by removing the events which
have classifier output value lower than the optimized cutoff value and using
the true positive ($S$) and false positive ($B$) events in~\cref{eq:significance}.
As noted in \cref{sec:point0}, the relative differences of the
significance values
obtained by different methods should be considered more
reliable than the absolute values.

\begin{figure}
    \centering
    \begin{subfigure}[t]{0.47\textwidth}
        \includegraphics[width=\textwidth]{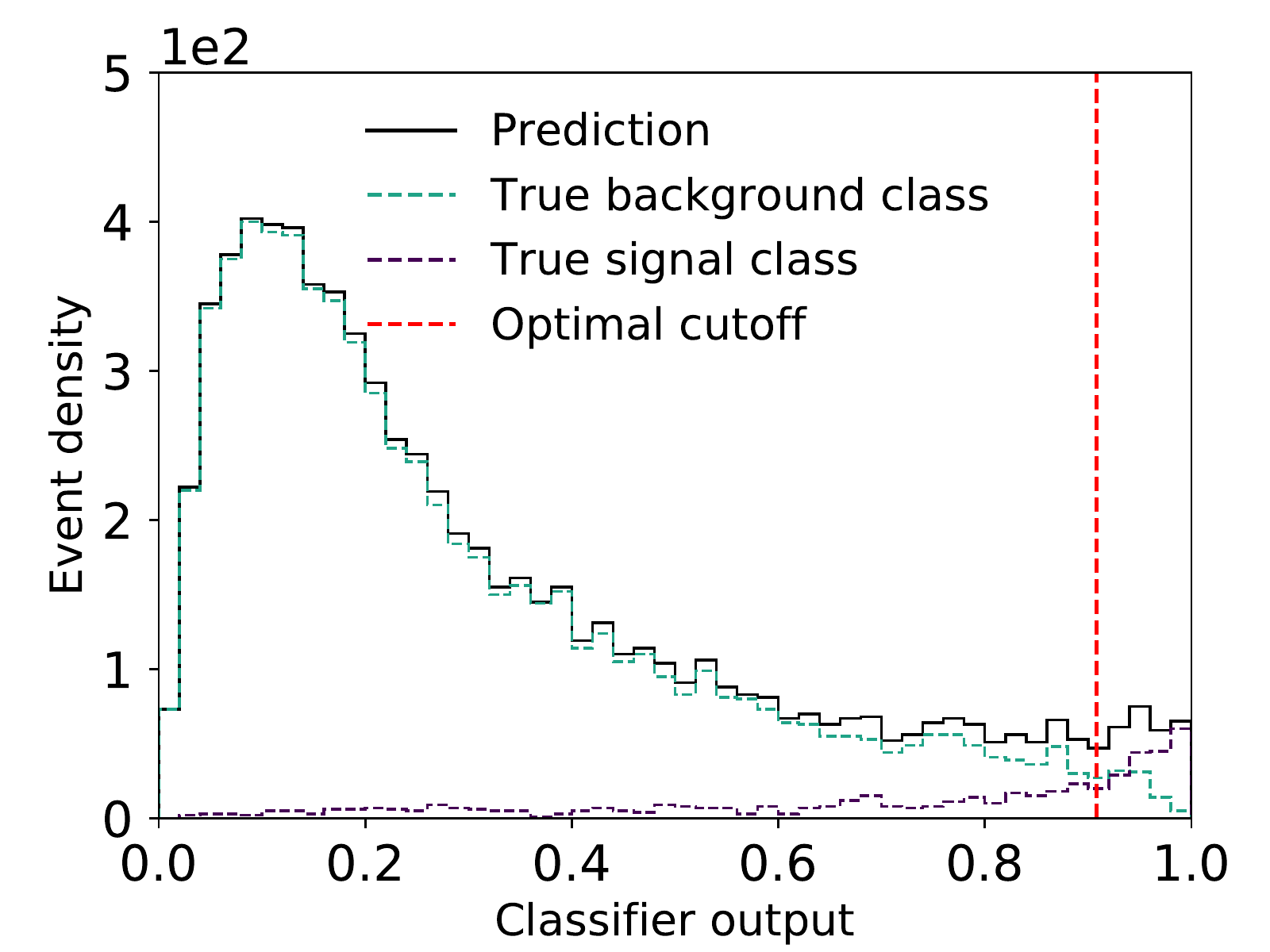}
        \caption{XGBoost with optimized cutoff at $0.9081$.}
    \end{subfigure}
    \quad
    \begin{subfigure}[t]{0.47\textwidth}
        \includegraphics[width=\textwidth]{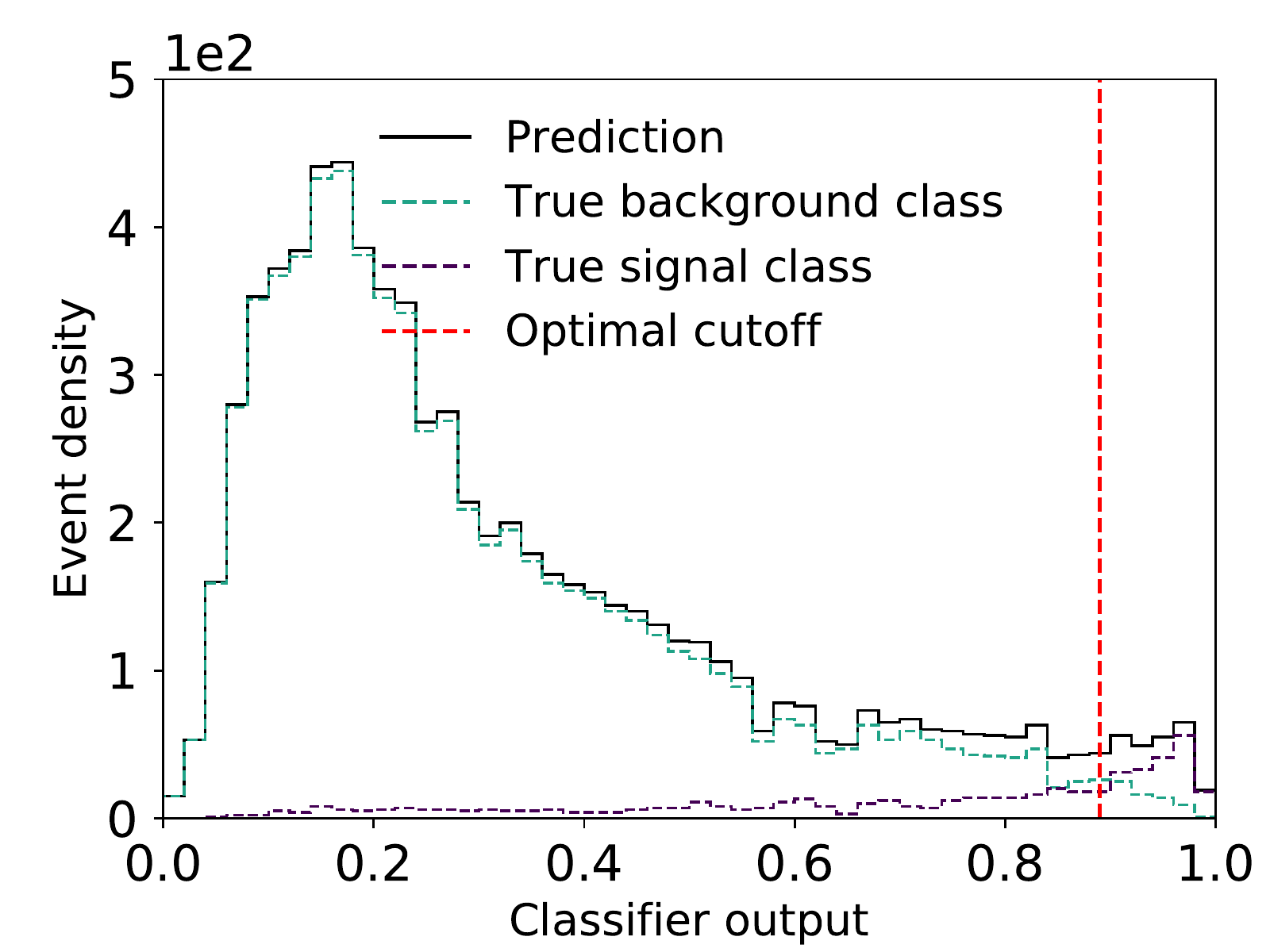}
        \caption{\ac{DNN} with optimized cutoff at $0.8896$.}
    \end{subfigure}
    \caption{\label{fig:classifier_cutoff_point12}Predictions of the \ac{ML} models on
    the test data set for Point 12, in a solid black line.  The true classes
    are shown in dashed green (\ac{SM} background) and dashed purple (signal), and
    the optimized cutoff in dashed red.  }
\end{figure}

For Point 12, the optimized cutoff values are found at $0.9081$ for the
XGBoost classifier and $0.8896$ for the \ac{DNN}. The classifier outputs
are shown in~\cref{fig:classifier_cutoff_point12}, with the optimized
cutoff values indicated. Applying these cutoff values to the Point 12
test data set using the two classifiers leaves $108$ background events (out of $6963$) 
and $198$ signal events (out of $522$), corresponding to $z=15.5$ for the
XGBoost classifier. 
The classifier is not able to correctly identify any of the four QCD events, which is not surprising as it was not trained on identifying such events.
For the \ac{DNN}, there remain $77$ background and $188$ signal events, corresponding to $z=16.7$. This is fewer signal events than for XGBoost, but the \ac{DNN} performs better on the background with $99\%$ correctly identified. This means that both classifiers outperform the cut analysis described in~\cref{sec:cut_analysis}, where the maximum significance achieved on Point~12 is $z=10.8$.

\begin{figure}
    \centering
    \begin{subfigure}[t]{0.47\textwidth}
        \includegraphics[width=\textwidth]{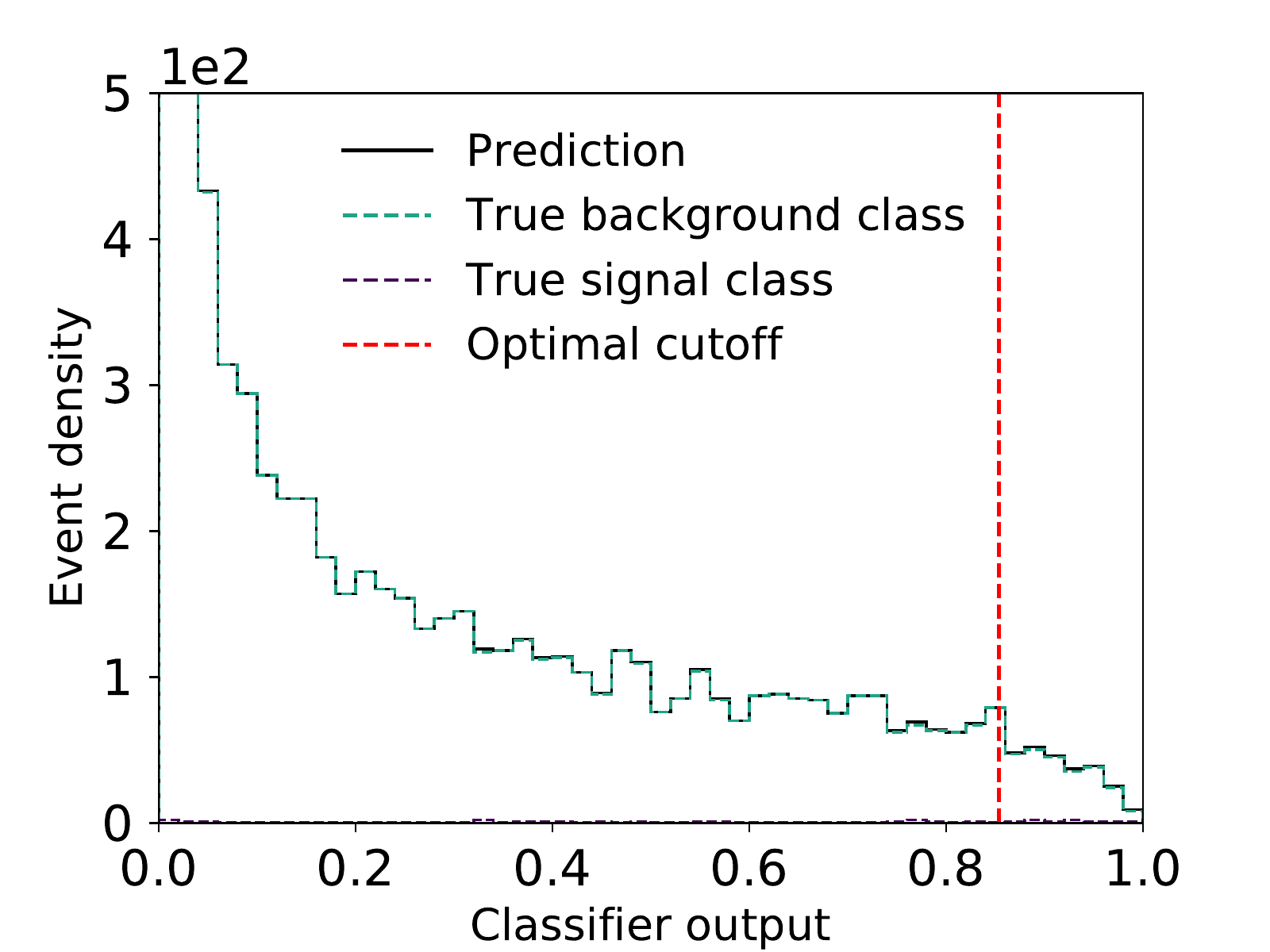}
        \caption{XGBoost with optimized cutoff at $0.8535$.}
    \end{subfigure}
    \quad
    \begin{subfigure}[t]{0.47\textwidth}
        \includegraphics[width=\textwidth]{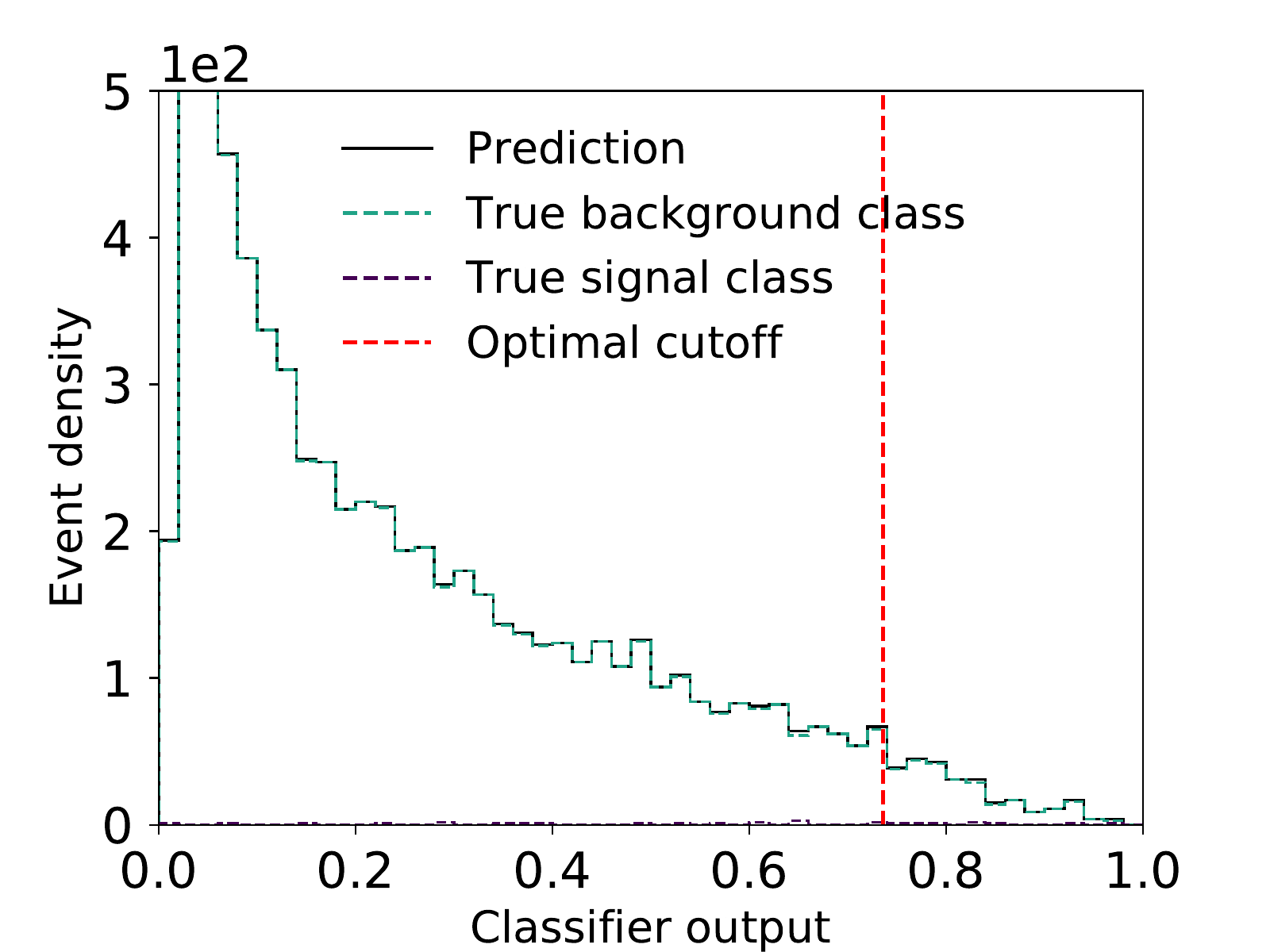}
        \caption{\ac{DNN} with optimized cutoff at $0.7356$.}
    \end{subfigure}
    \caption{\label{fig:classifier_cutoff_point0} Predictions of the \ac{ML} models on
    the test data set for Point 0, in a solid black line.  The true classes
    are shown in dashed green (\ac{SM} background) and dashed purple (signal), and
    the optimized cutoff in dashed red.
    }
\end{figure}

For Point 0, the optimized cutoff values are found at $0.8535$ for the XGBoost
classifier and $0.7356$ for the \ac{DNN}.  The classifier outputs and optimized
cutoff values are shown in~\cref{fig:classifier_cutoff_point0}.  For the
Point~0 test data set these cutoff values lead to $275$ background and $10$
signal events (out of $24$), corresponding to $z=0.57$, for the XGBoost classifier and to
$270$ background and $10$ signal events, corresponding to $z=0.63$, for the
\ac{DNN}.  Thus, also for this parameter space
point both classifiers outperform the cut analysis, whose maximum significance
is $z=0.3$ here.  However, this improvement is not sufficient for detection
with the considered luminosity.

The classifier for Point 12 is also applied to the other points, i.e., Points
0, 13--16, 20, 30, 40, and 50, using the same method.  We use the optimized
cutoff value from Point 12 on the classifier output to select signal and
background events. Since some of the points contain very few signal events this
could lead to an over- or underestimation of the significance. We therefore
scale up the number of test events and then scale down again $S$ with the same
factor. The significances using both the XGBoost and DNN classifiers can be
found in~\cref{tab:point_fits} in the columns $z_\text{XGB-cut}$ and
$z_\text{DNN-cut}$, respectively.

\subsection{Estimating the signal mixture parameter}\label{sec:mle}
While it is encouraging that \ac{ML} models can outperform our analysis
from~\cref{sec:cut_analysis}, this is on the one hand not a novelty, and on the
other hand of limited practical usefulness, given that we do not know which
combination of \ac{SUSY} parameters is realized in nature.  One way of
approaching this is by identifying regions in the parameter space which share
similar signal kinematics and train an \ac{ML} classifier on the expected
signal and the \ac{SM} background. Such a classifier should then show robust
performance within such a region, by recognizing the familiar (and unchanging)
\ac{SM} background and partly recognizing the signal, which changes only a
little bit within the region and in any case resembles the background less than
the similar signal from the training data.  However, not only the kinematics of
the signal change throughout the parameter space but also the
signal-to-background ratio, upon which our optimized cutoff value depends.
Ideally, a detection method should be independent of the signal
\textit{admixture}, and so we reformulate our problem as a mixture parameter
estimation task in the following section.  Although only XGBoost and a \ac{DNN}
are considered here, the method can be used for any classifier which maps the
input features to continuous values.

The distribution of signal and background data can be expressed as a simple
mixture model
\begin{equation}
    p = \alpha p_{s} + (1-\alpha)p_{b} \,,
\end{equation}
where $p_{b/s}$ denote the probability density\footnote{Strictly speaking,
probability \textit{mass} functions.} functions for \textit{b}ackground and
\textit{s}ignal, respectively, and $\alpha$ represents the signal mixture
parameter. 

\begin{figure}
    \centering
    \begin{subfigure}[t]{0.47\textwidth}
        \includegraphics[width=\textwidth]{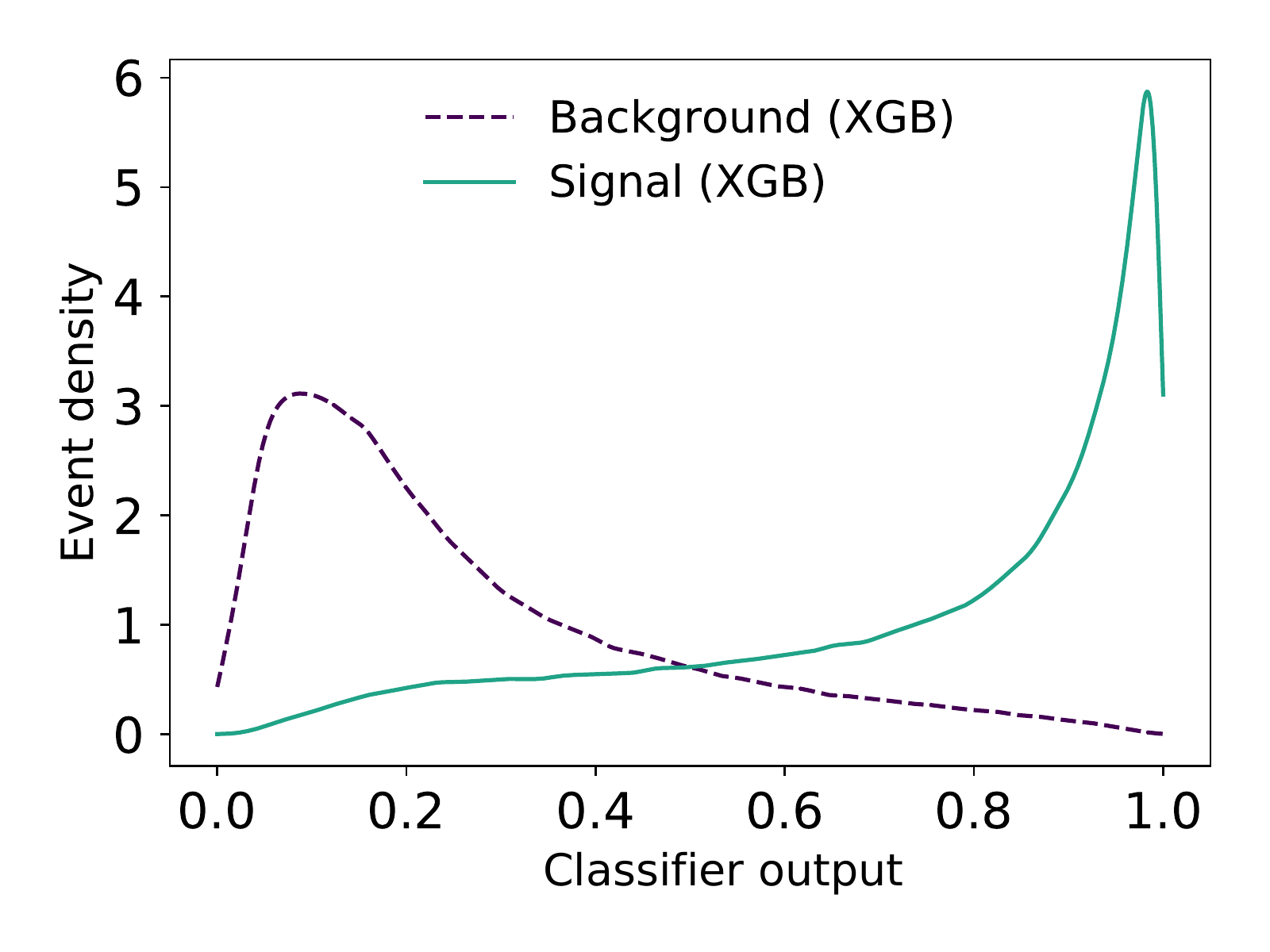}\vspace{-1.5ex}
        \caption{\label{fig:templates_xgb}}
    \end{subfigure}
    \quad
    \begin{subfigure}[t]{0.47\textwidth}
        \includegraphics[width=\textwidth]{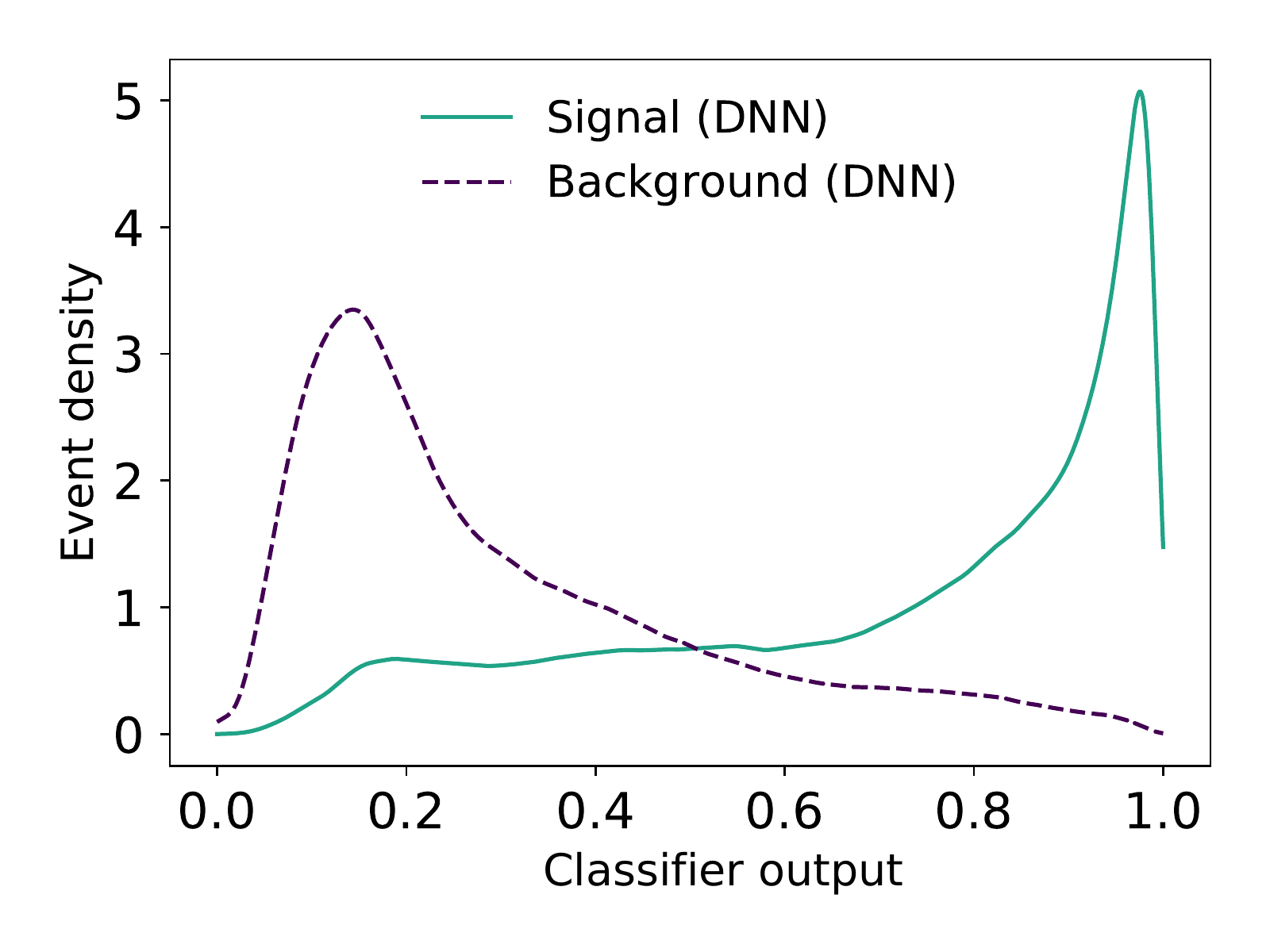}\vspace{-1.5ex}
        \caption{\label{fig:templates_dnn}}
    \end{subfigure}
    \caption{\label{fig:templates}Signal and background class templates
    created using (\protect\subref{fig:templates_xgb}) the XGBoost and 
    (\protect\subref{fig:templates_dnn}) \ac{DNN} classifier, trained on Point~12 data (see~\cref{sec:point12}).}
\end{figure}

We can estimate the probability densities $p$ by constructing class templates
from the trained classifiers as follows. We let the classifiers predict on one
data set containing only background events, which yields $p_b$, and on
one data set containing only signal events, which yields $p_s$.
We use kernel density estimation with a
Gaussian kernel, renormalized on the edges to properly cover the area around
$0$ and $1$, to have a continuous representation of the templates. 
From the training data we set aside $400\,000$ events and use 
these to construct the templates, which are shown for Point~12 in~\cref{fig:templates}.
The optimal number of events to use for template creation was found by testing.
In our approach, uncertainty due to the amount of Monte Carlo events
available arises
in two places -- in training the classifiers and in constructing the templates. 
For the former, the relationship between prediction uncertainty and number of 
training events is not trivial, as it also depends on the classifier's complexity
and the training procedure itself, including issues such as under- or overfitting.
One can assume, however, that the uncertainty for a given classifier will decrease
with increasing number of training events, until a plateau is reached, where additional
improvement would require a more complex model (i.e., the classifier is underfitting).
Uncertainty in the template shape is more directly related to the number of
Monte Carlo events used, but complicated by the fact that the tail of the background
template distribution (shown in~\cref{fig:templates}) is the most important
for the signal mixture estimation.

Next, we perform the admixture estimation by letting the classifiers predict on a set of previously unseen data, and fit the templates in~\cref{fig:templates} to the corresponding classifier outputs using an unbinned maximum likelihood fit. The fit returns the estimated admixture of the two models described by a background and signal template, respectively, that maximizes the likelihood of the classifier's output for the given data set. We use $\hat{\alpha}$ to denote the method's estimate of the mixture parameter.

\begin{figure}
    \centering
    \begin{subfigure}[t]{0.3\textwidth}
        \includegraphics[width=\textwidth]{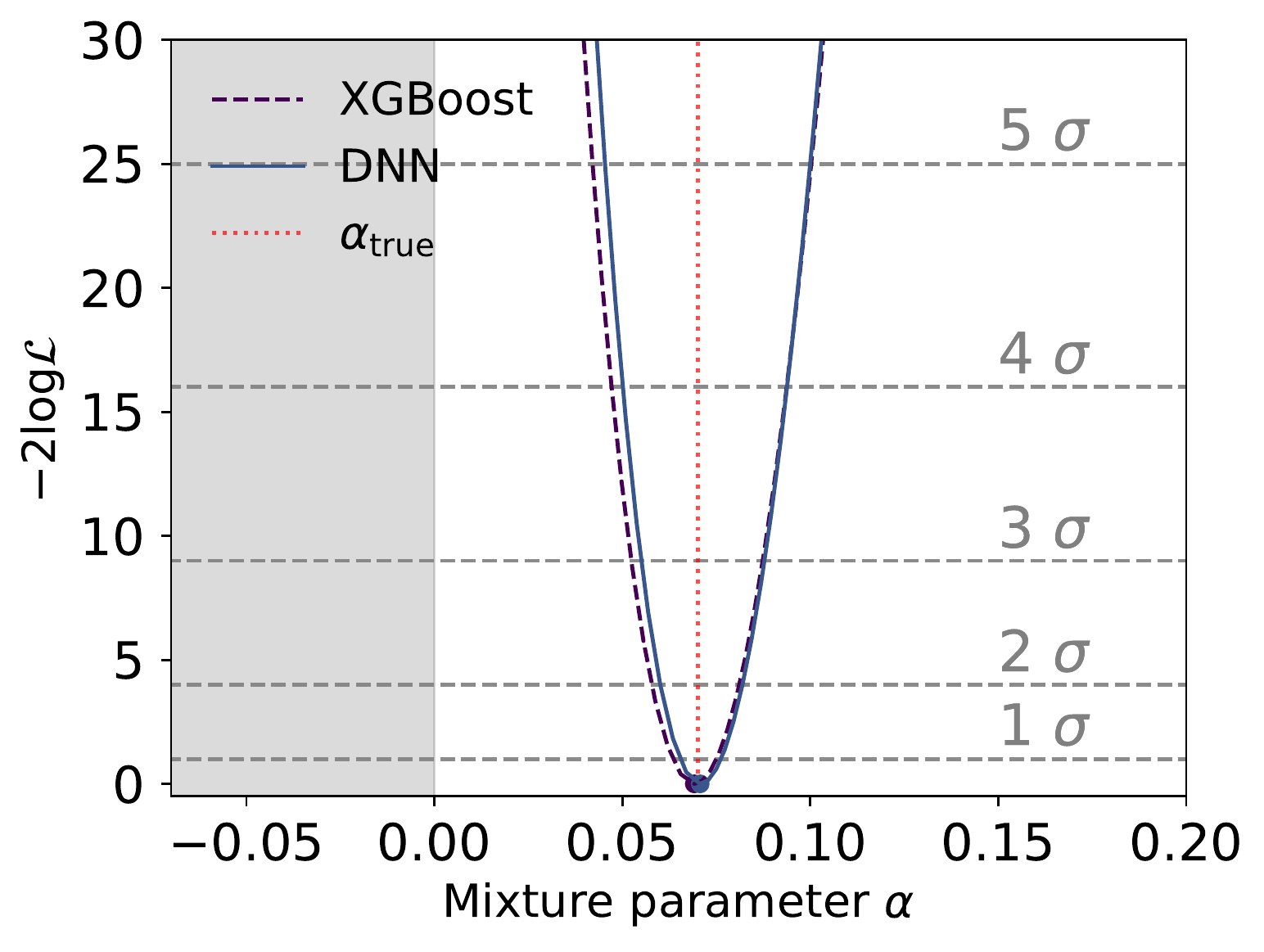}
        \caption{\label{fig:ll_fits_nn_p12}
        Point 12:
        $\alpha=0.070$,\\
        $\hat\alpha_{\text{XGB}}=0.069$, 
        $\hat\alpha_{\text{DNN}}=0.071$.}
    \end{subfigure}
    \begin{subfigure}[t]{0.3\textwidth}
        \includegraphics[width=\textwidth]{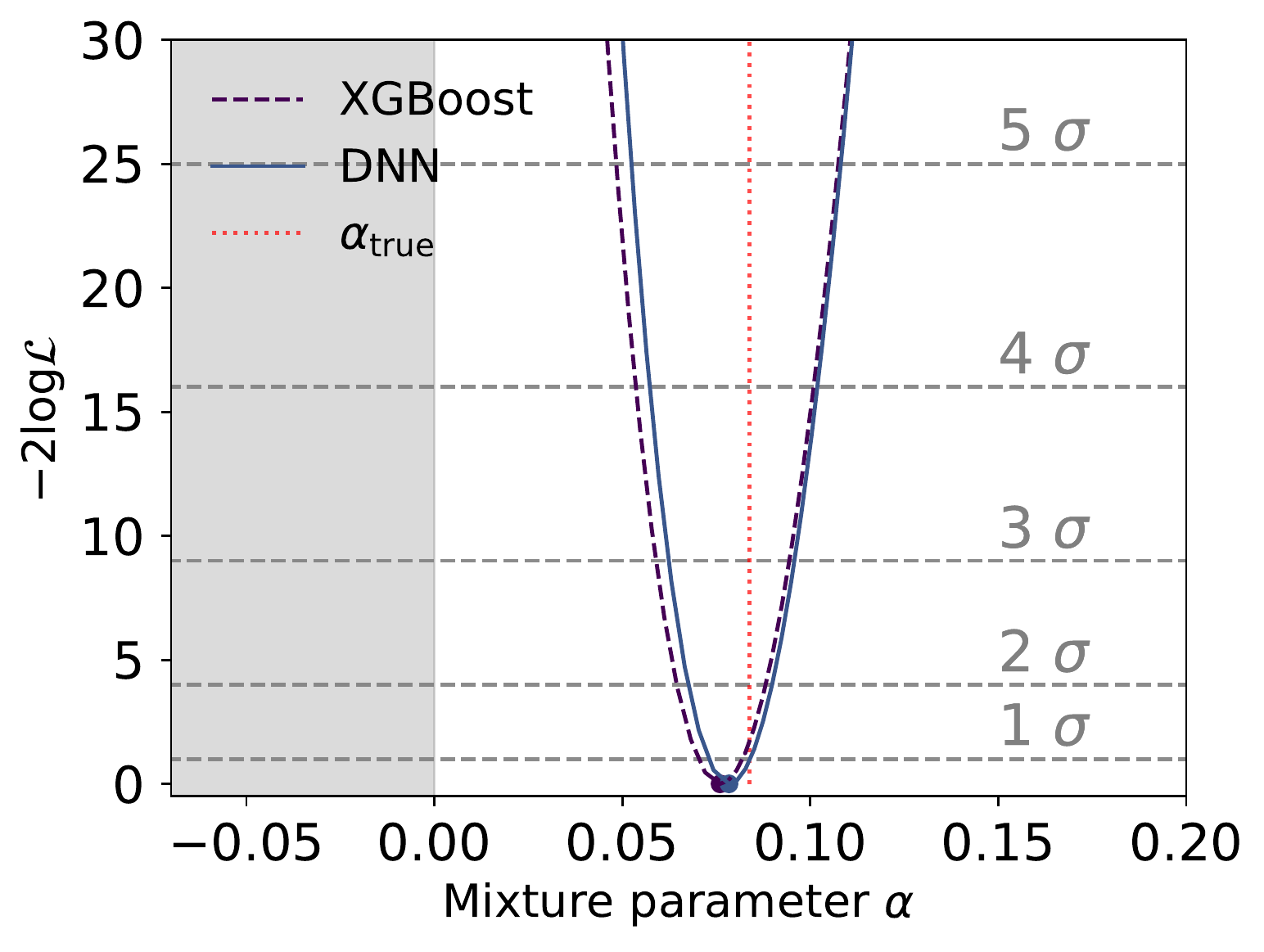}
        \caption{
        Point 13:
        $\alpha=0.084$,\\
        $\hat\alpha_{\text{XGB}}=0.076$, 
        $\hat\alpha_{\text{DNN}}=0.078$.}
    \end{subfigure}
    \begin{subfigure}[t]{0.3\textwidth}
        \includegraphics[width=\textwidth]{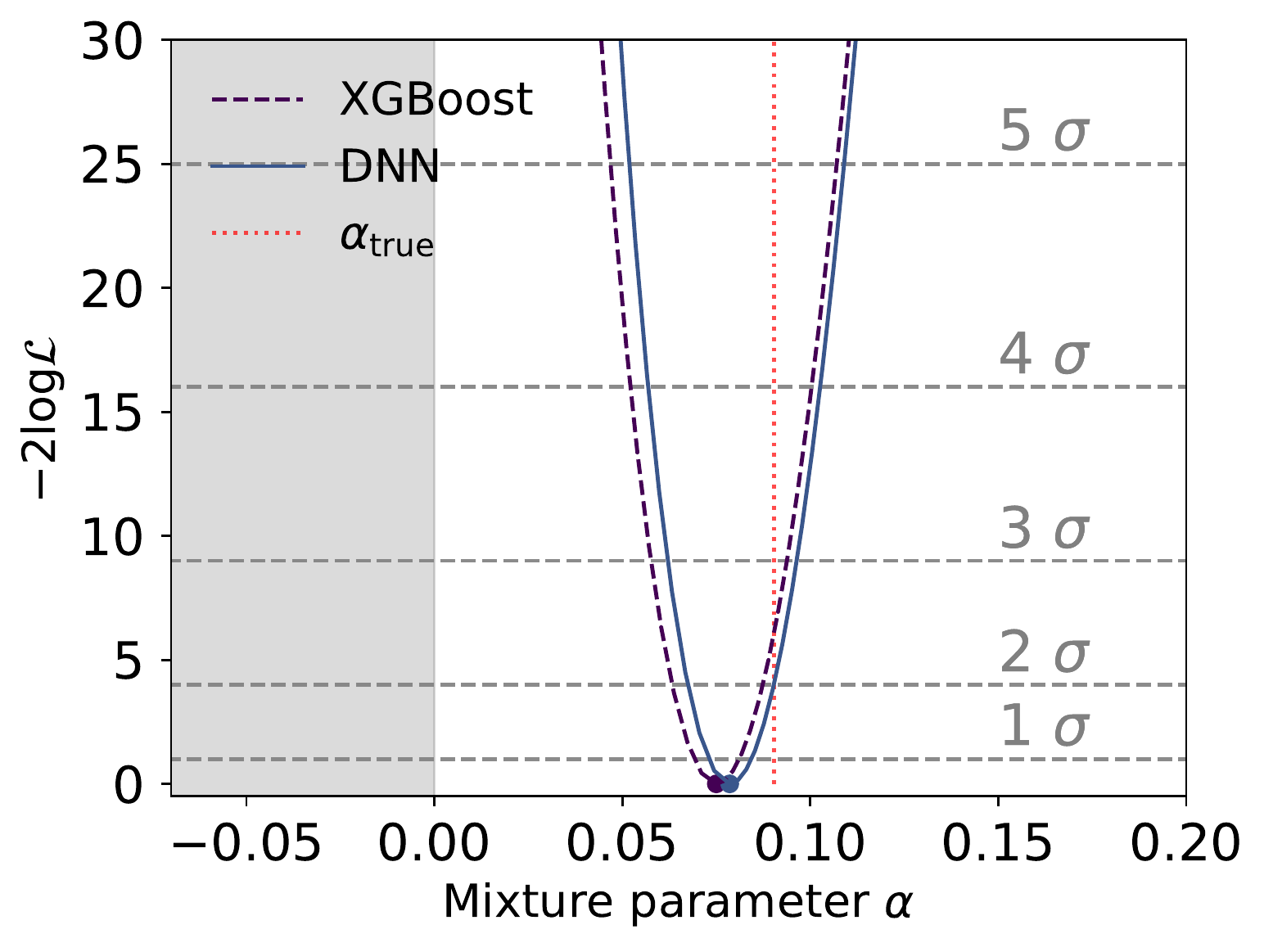}
        \caption{
        Point 14:
        $\alpha=0.090$,\\
        $\hat\alpha_{\text{XGB}}=0.075$, 
        $\hat\alpha_{\text{DNN}}=0.079$.}
    \end{subfigure}
    \begin{subfigure}[t]{0.3\textwidth}
        \includegraphics[width=\textwidth]{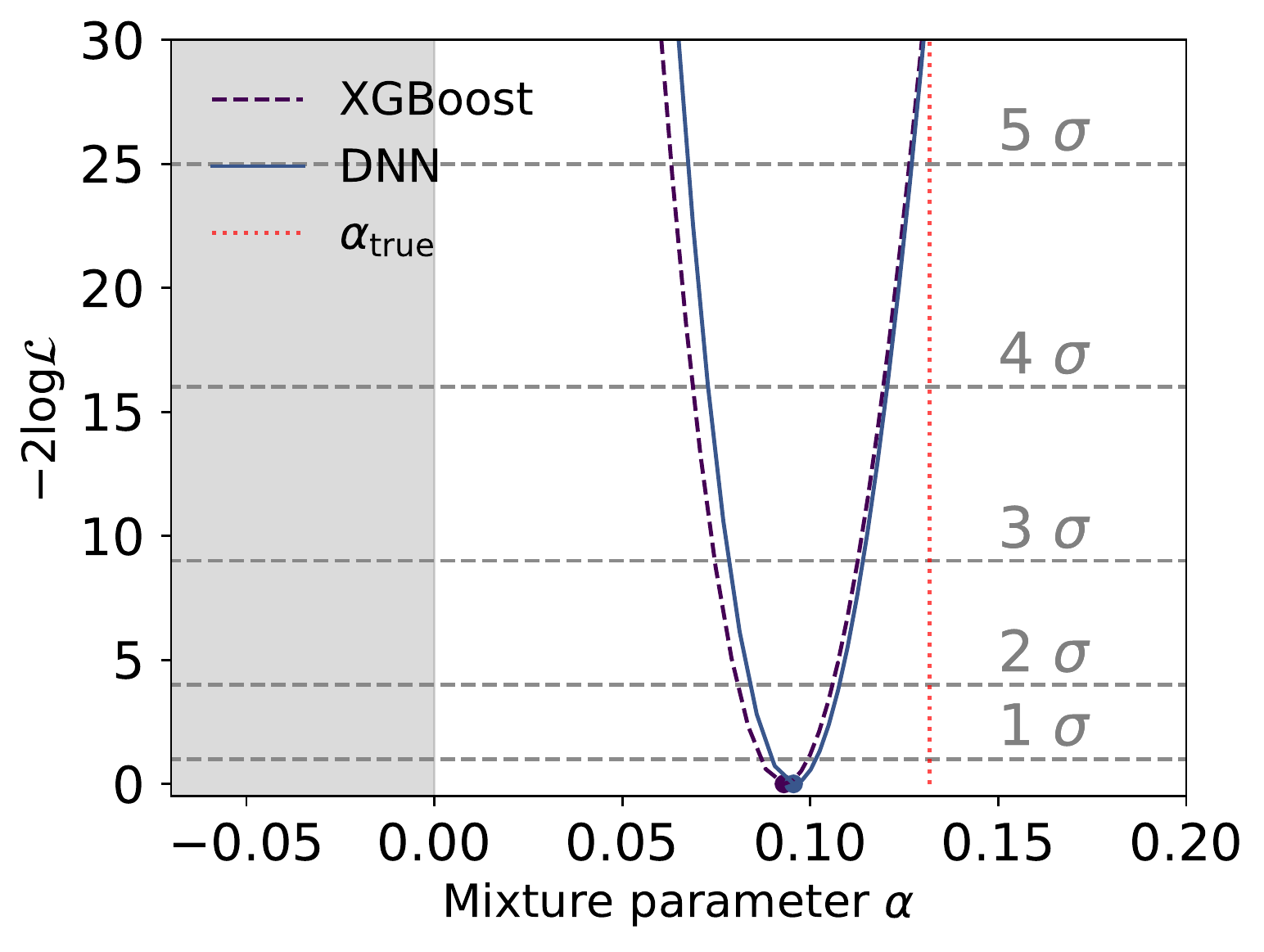}
        \caption{
        Point 15:
        $\alpha=0.132$,\\
        $\hat\alpha_{\text{XGB}}=0.093$, 
        $\hat\alpha_{\text{NNN}}=0.096$.}
    \end{subfigure}
    \begin{subfigure}[t]{0.3\textwidth}
        \includegraphics[width=\textwidth]{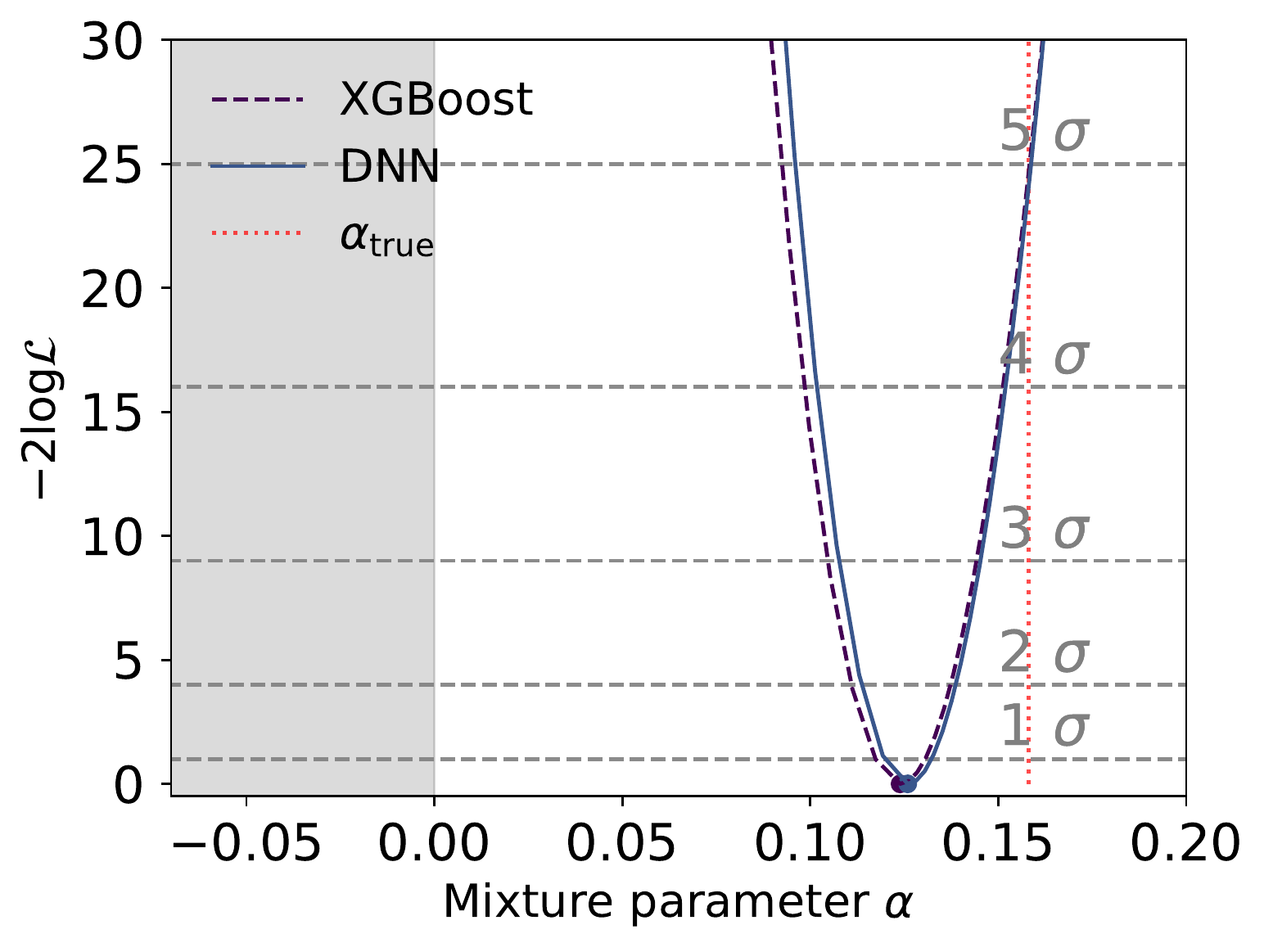}
        \caption{
        Point 16:
        $\alpha=0.158$,\\
        $\hat\alpha_{\text{XGB}}=0.124$,
        $\hat\alpha_{\text{DNN}}=0.126$.}
    \end{subfigure}
    \begin{subfigure}[t]{0.3\textwidth}
        \includegraphics[width=\textwidth]{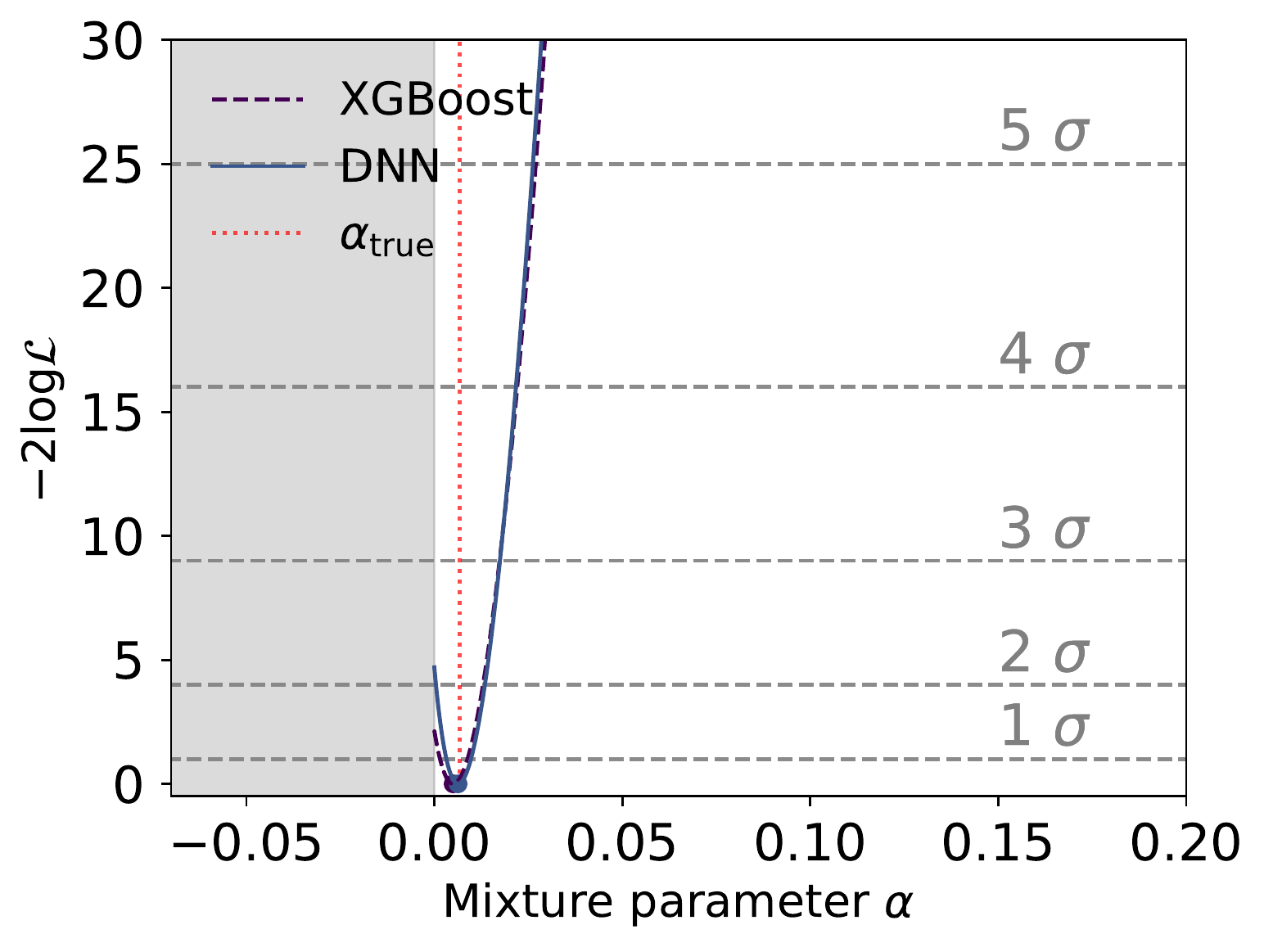}
        \caption{
        Point 20:
        $\alpha=0.007$,\\
        $\hat\alpha_{\text{XGB}}=0.005$, 
        $\hat\alpha_{\text{DNN}}=0.006$.}
    \end{subfigure}
    \begin{subfigure}[t]{0.3\textwidth}
        \includegraphics[width=\textwidth]{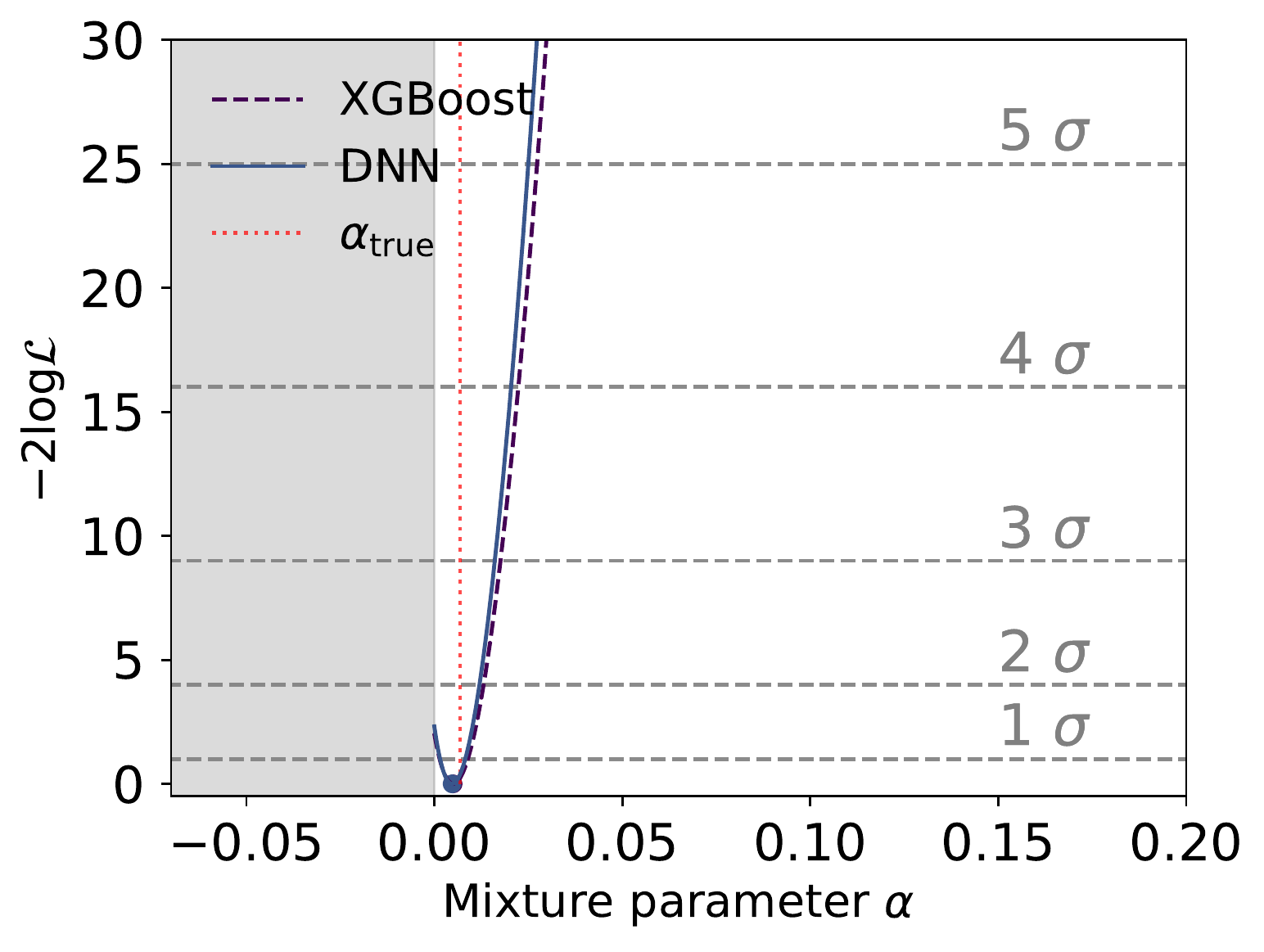}
        \caption{
        Point 30:
        $\alpha=0.007$, \\
        $\hat\alpha_{\text{XGB}}=0.005$, 
        $\hat\alpha_{\text{DNN}}=0.005$.}
    \end{subfigure}
    \begin{subfigure}[t]{0.3\textwidth}
        \includegraphics[width=\textwidth]{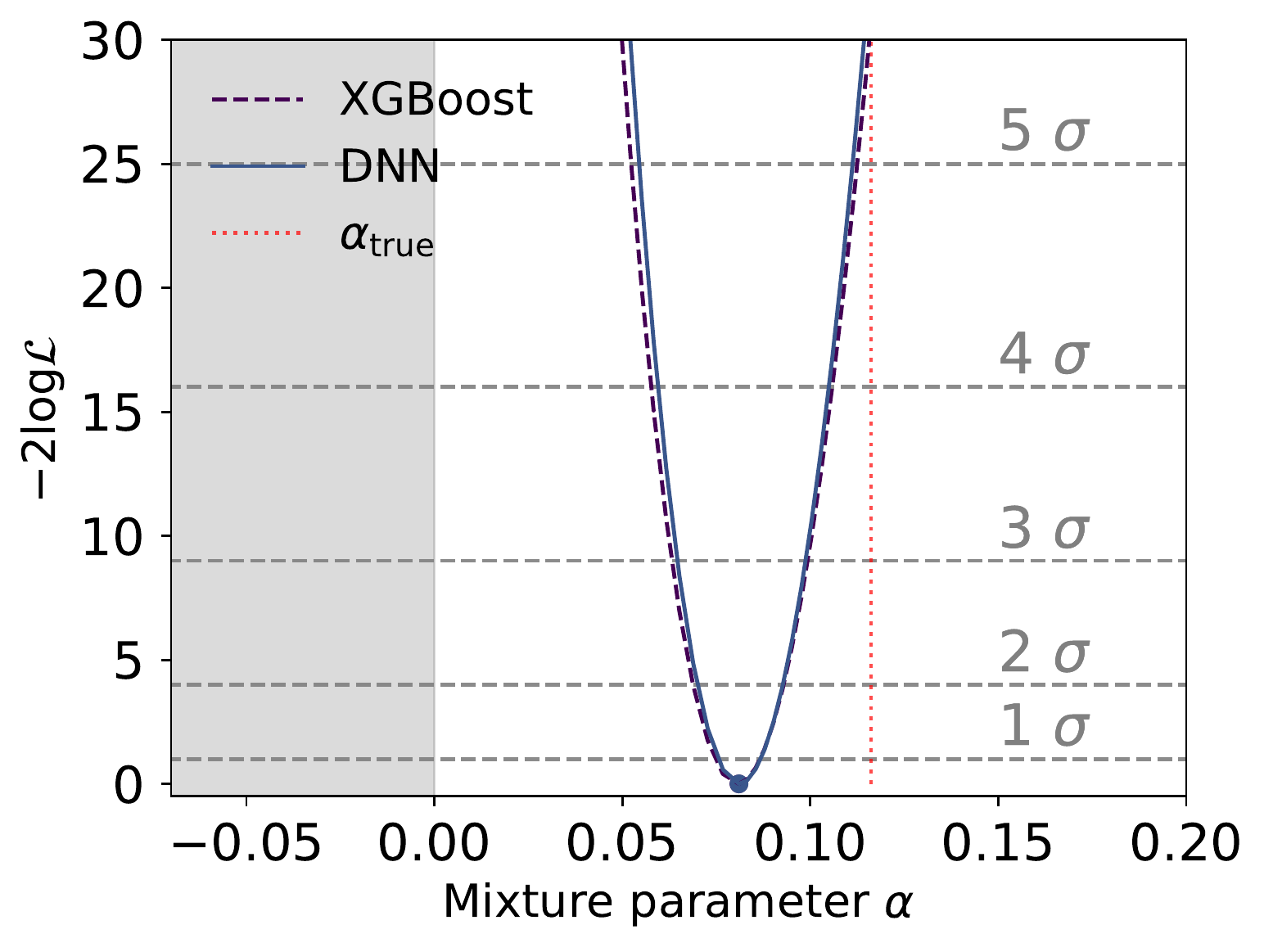}
        \caption{
        Point 40:
        $\alpha=0.116$, \\
        $\hat\alpha_{\text{XGB}}=0.081$, 
        $\hat\alpha_{\text{DNN}}=0.081$.}
    \end{subfigure}
    \begin{subfigure}[t]{0.3\textwidth}
        \includegraphics[width=\textwidth]{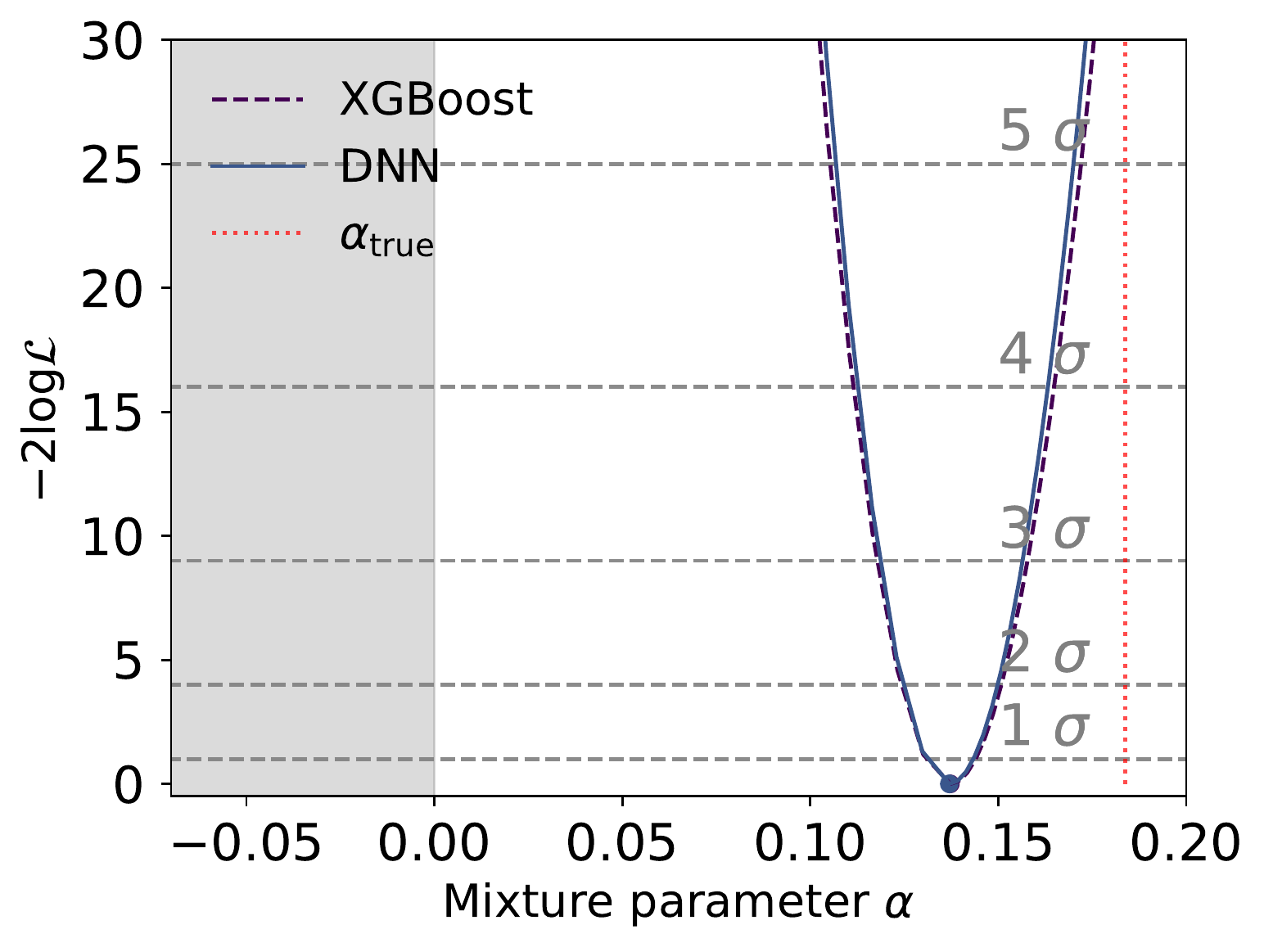}
        \caption{
        Point 50:
        $\alpha=0.184$, \\
        $\hat\alpha_{\text{XGB}}=0.137$, 
        $\hat\alpha_{\text{DNN}}=0.137$.}
    \end{subfigure}
    \begin{subfigure}[t]{0.3\textwidth}
        \includegraphics[width=\textwidth]{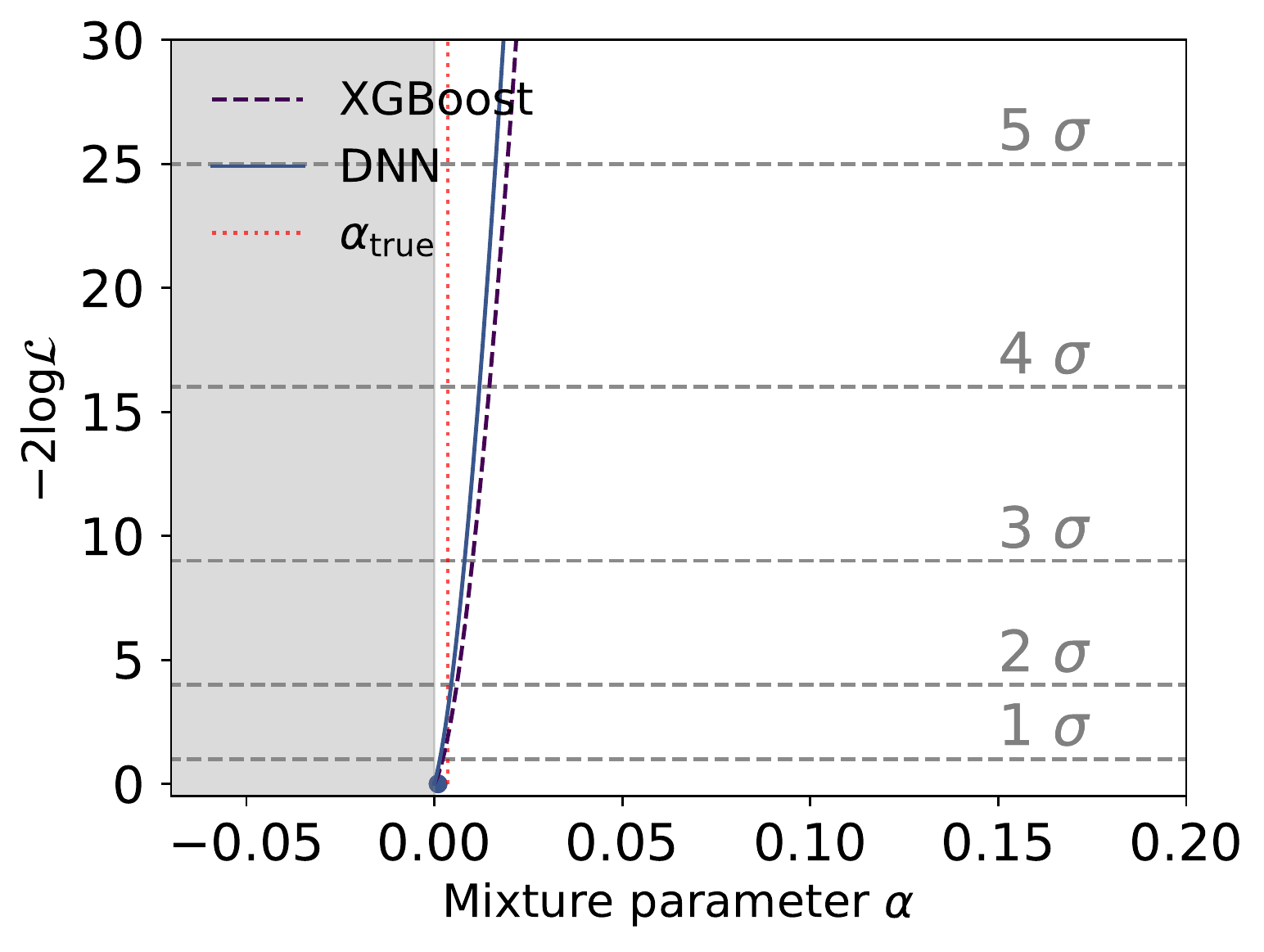}
        \caption{
        Point 0:
        $\alpha=0.004$, \\
        $\hat\alpha_{\text{XGB}}=0.001$, 
        $\hat\alpha_{\text{DNN}}=0.001$.}
    \end{subfigure}
    \caption{\label{fig:ll_fits_nn}
    Maximum likelihood fits to the different parameter points, using the
    XGBoost (dashed red) and \ac{DNN} (solid blue) models. 
    The true mixture parameter $\alpha$ and the best-fit mixture parameters
    $\hat\alpha$ are given in the subcaptions, and listed in with their $95\%$
    confidence intervals in~\cref{tab:point_fits}.
    } 
\end{figure}

The test data set for Point 12 (cf.~\cref{sec:point12}) has a signal mixture
parameter of $\alpha_{\textrm{true}} = 0.07$, which is challengingly small.
Using the described procedures, we obtain best-fit estimates of
$\hat\alpha_{\text{xgb}} = 0.069 \pm 0.011$ and $\hat\alpha_{\text{DNN}} = 0.071 \pm 0.011$,
respectively, 
where the uncertainties are statistical. Theoretical uncertainties on
the fit results are discussed in \cref{sec:signal_variation_comparison}.
The corresponding log-likelihood curves with the $n$ sigma
regions indicated are shown in~\cref{fig:ll_fits_nn_p12}.

Nature may choose any \ac{SUSY} parameter point, resulting in some signal
mixture parameter $\alpha_{\mathrm{true}}$, which is of course unknown to us.
The determination of this mixture parameter from potential data collected in a
particle collider will aid primarily in signal detection and furthermore in
fixing the parameters of the underlying \ac{SUSY} model.  Since there is
an uncountable number of different \ac{SUSY} parameter points,
we want to investigate whether a classifier trained on kinematics representing
one parameter point can generalize to, i.e., estimate the signal mixture
parameter of, other parameter points featuring the same type of signal, but
different kinematics.
In order to test this, we train a classifier on Point 12
and use this classifier to predict on the different
parameter points listed in~\cref{tab:parampoints_yields,tab:parampoints_details}.
Based on how these different points are distributed, we claim to have a
representative sample for investigating how well the method generalizes as the
\ac{SUSY} input parameters change.
To estimate the different points' signal mixture parameters, we perform an
unbinned maximum likelihood fit to the classifier output for each point to
determine the best-fit estimate mixture parameters $\hat{\alpha}$. The results
are listed in~\cref{tab:point_fits}. We also show the log-likelihood curves
indicating the $n$ sigma region in~\cref{fig:ll_fits_nn}, along with the
best-fit values. 

\begin{table}
    \centering
    \small
    \caption{Estimated mixture parameters for the
    different points using the two classifiers described in the text and the
    corresponding discovery significances.
    Both \ac{ML} classifiers are trained on Point~12.}
    \begin{tabular}{lSllSSSSS}
        & $\alpha_{\mathrm{true}}$ & $\hat{\alpha}_{\text{XGB}}$  &
        $\hat{\alpha}_{\text{DNN}}$ & $z_{\text{XGB}}$ & $z_{\text{DNN}}$
        & $z_{\text{XGB-cut}}$ & $z_{\text{DNN-cut}}$ &  $z_{\text{Cut\&count}}$ \\
        \midrule
        Point 12 & 0.070  & $0.069 \pm 0.011$   & $0.071 \pm 0.011$ & 15.7 & 18.5 & 15.5 & 16.7 & 10.8\\
        Point 13 & 0.084  & $0.076 \pm 0.011$   & $0.078 \pm 0.012$ & 17.2 & 20.3 & 16.6 & 17.3 & 12.7\\
        Point 14 & 0.090  & $0.075 \pm 0.011$   & $0.079 \pm 0.012$ & 16.3 & 19.2 & 15.9 & 16.4 & 11.0\\
        Point 15 & 0.132  & $0.093 \pm 0.012$   & $0.096 \pm 0.012$ & 19.4 & 22.9 & 19.9 & 20.1 & 13.6\\
        Point 16 & 0.158  & $0.124 \pm 0.013$   & $0.126 \pm 0.013$ & 25.9 & 29.7 & 22.7 & 22.6 & 19.0\\
        Point 20 & 0.007  & $0.005 \pm 0.008$   & $0.006 \pm 0.006$ & 1.5  & 2.2  & 1.5  & 1.6  & 1.2\\
        Point 30 & 0.007  & $0.005 \pm 0.009$   & $0.005 \pm 0.007$ & 1.4  & 1.5  & 1.9  & 2.1  & 1.6\\
        Point 40 & 0.116  & $0.081 \pm 0.012$   & $0.081 \pm 0.012$ & 17.8 & 20.2 & 16.5 & 16.8 & 11.2\\
        Point 50 & 0.184  & $0.137 \pm 0.014$   & $0.137 \pm 0.014$ & 28.7 & 32.5 & 28.2 & 28.7 & 20.5\\
        Point 0  & 0.004  & $0.001 \pm 0.006$   & $0.001 \pm 0.004$ & 0.0  & 0.0  & 0.2  & 0.2  & 0.1\\
        \bottomrule
    \end{tabular}
    \label{tab:point_fits}
\end{table}

\subsection{Shapley decomposition of labels and model predictions}
Before comparing the two \ac{ML} based approaches, we address the well known challenge that large \ac{ML} models, such as \acp{DNN}, have a black-box nature. Although we have access to the input data and all the tuned parameters, this does not necessarily tell us to which features or combinations of features the model assigns importance. This is the central issue in the field of \ac{XAI}, and various methods have been proposed to address this challenge. One of these is the Shapley decomposition, a solution concept from cooperative game theory first introduced in~\cite{shapley}, which has become popular in the \ac{XAI} literature~\cite{NIPS2017_7062,[14],[16],[34],[36],[43],[F15]}. 
Recent
studies~\cite{grojean2021,cornell2022boosted,WALKER2022137255,alasfar2022machine,bhattacherjee2022boosted}
also used a variety of methods based on Shapley decomposition to explain their \ac{ML} models, and the importance of explanation methods for interpreting the output of \ac{ML} models used in particle physics is discussed in the overview provided in~\cite{grojean2022lessons}.

To understand the importance our \ac{ML} classifiers give the different features in~\cref{tab:input_feat}, and whether our models accurately capture the dependence structure between these features and the labels, we calculate the Shapley decomposition among the features and the labels, denoted \ac{ADL}, as well as the Shapley decomposition of the features and the classifiers' predictions, denoted \ac{ADP}, using as utility function the distance correlation~\cite{dcor}, as detailed in~\cite{sunnies}.  

\begin{figure}
    \centering
    \begin{subfigure}[t]{0.7\textwidth}
        \includegraphics[width=\textwidth]{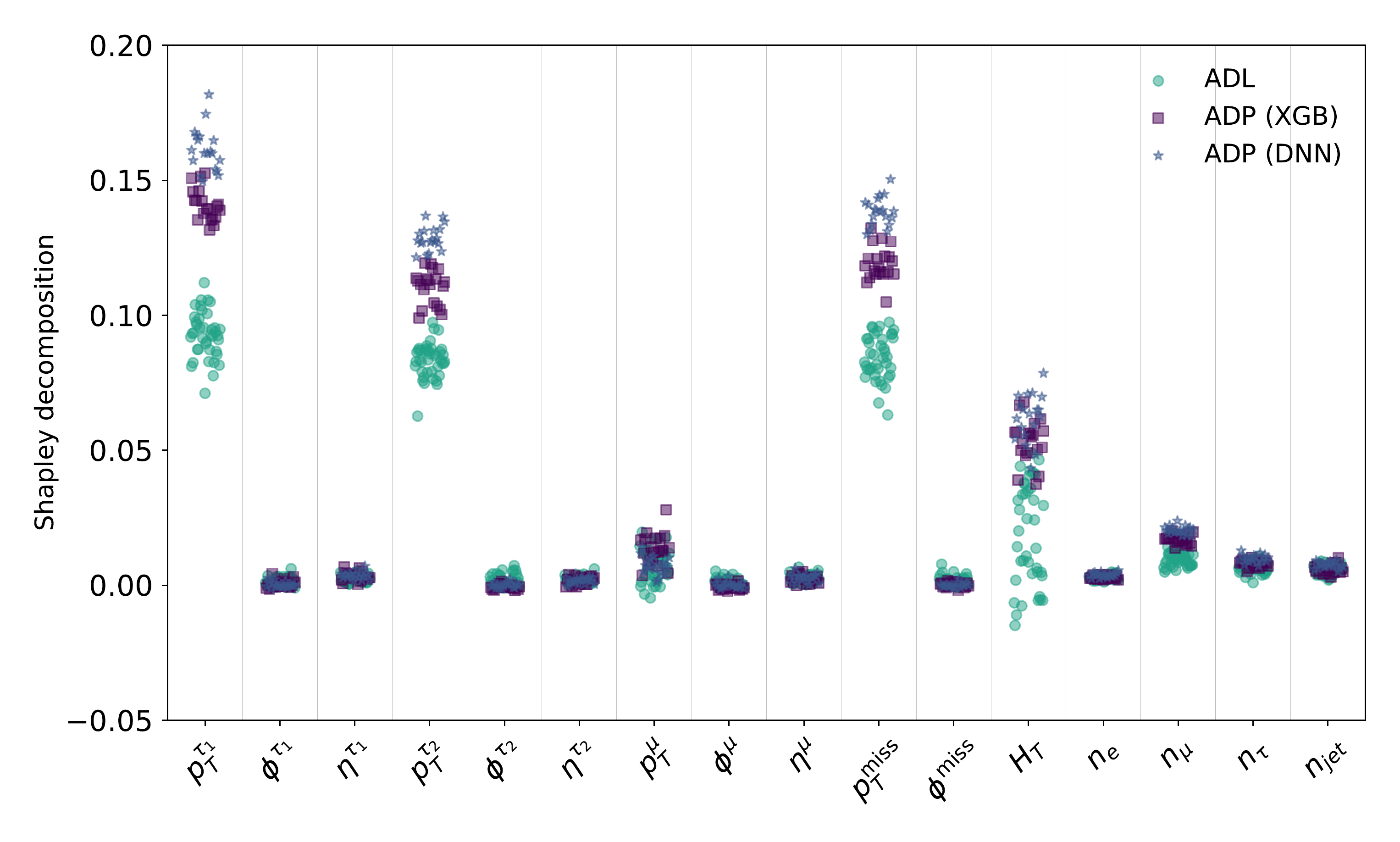}\vspace{-1ex}
        \caption{\label{fig:shapley_barbells} }
    \end{subfigure}
    \begin{subfigure}[t]{0.7\textwidth}
    \medskip
    \includegraphics[width=\textwidth]{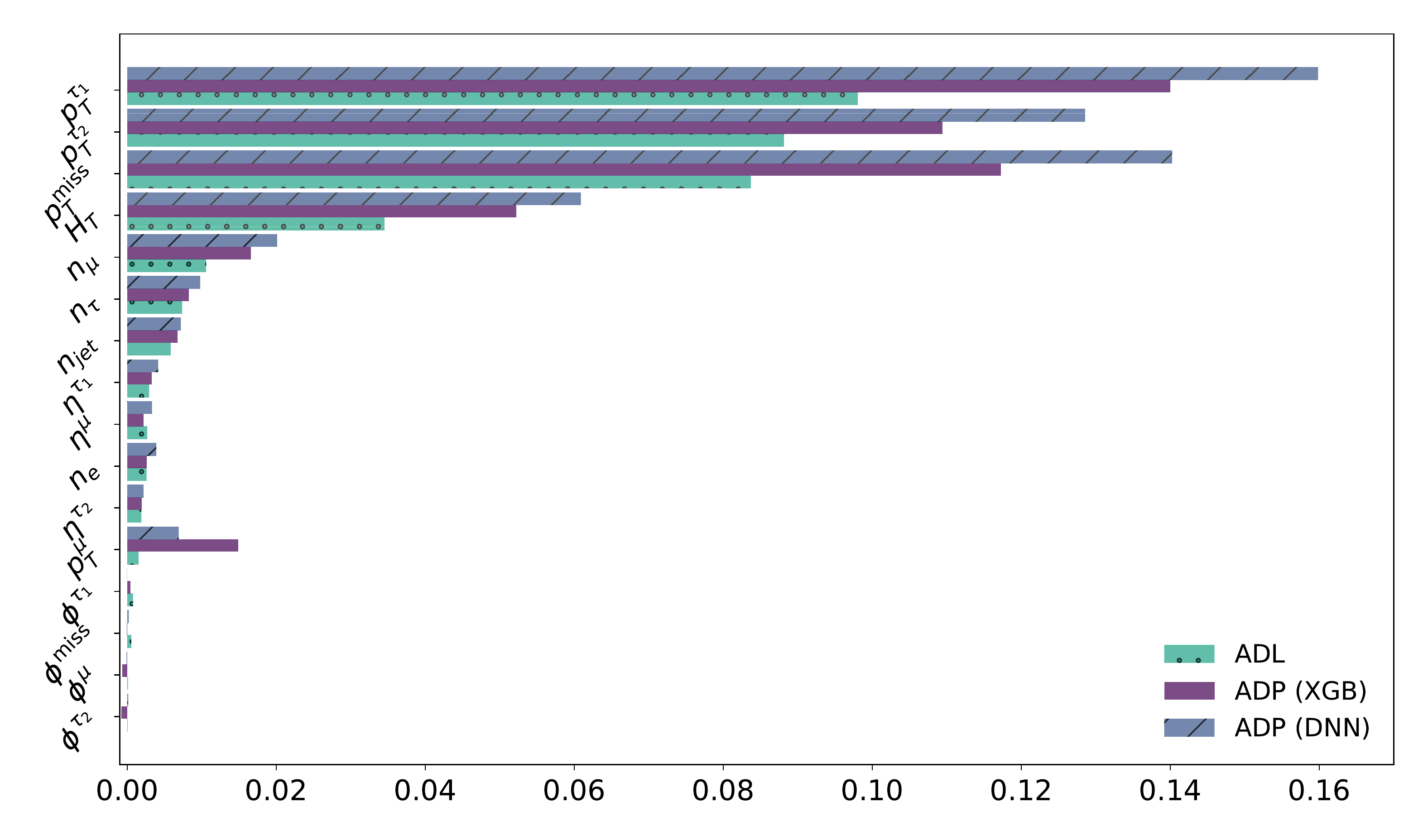}\vspace{-1ex}
        \caption{\label{fig:shapley_bars}}
    \end{subfigure}
    \caption{\label{fig:shapley}%
    (\protect\subref{fig:shapley_barbells})
    \acp{ADL} (green circles) and \acp{ADP} for the XGBoost (purple squares)
    and the \ac{DNN} (blue stars) models, using the distance correlation as
    utility function for the Shapley value.
    (\protect\subref{fig:shapley_bars})
    The average \ac{ADL} (dotted green) and \ac{ADP} for the XGBoost (purple)
    and the \ac{DNN} (hashed blue) models, per feature as indicated on the
    $y$-axis, in increasing order sorted according to the \ac{ADL}.
    }
\end{figure}

We can calculate point estimates for the \ac{ADL} and \ac{ADP} Shapley values
using our test data.
For calculating confidence intervals for the Shapley values, the
asymptotic distribution of the utility function must be known, and have a
finite variance.  As noted in~\cite{dcor}, section 2.4, the distance
correlation can be written as a V-statistic with a degenerate kernel, which
implies that the asymptotic distribution is not normal. Hence, we are not able
to calculate confidence intervals exactly for this utility
function\footnote{Using as utility function the more traditional coefficient of
determination, the $R^2$, the confidence intervals can be
calculated~\cite{shapley_cis}. However, the $R^2$ captures only linear
correlations, which we know are not sufficient for our problem.}, but as
explained in~\cite{Huettner:2012aa}, we can quantify the variability in the
Shapley values via bootstrap.  We do this by resampling $2000$ events ten times
from a data set containing $9000$ events equally distributed between signal and
background.  \Cref{fig:shapley_barbells} shows the \ac{ADL} and \ac{ADP} values
obtained by doing this for both the XGBoost and the \ac{DNN} model. The gray
lines are drawn between the average value in each group,
and~\cref{fig:shapley_bars} shows the average values per feature for the
\acp{ADL}, and the \acp{ADP} for both models.  The model predictions show an
\textit{overall} stronger correlation with the input features than the labels,
which is expected as long as the model performs reasonably well, since the
predictions are then a function of the features with no irreducible error term.

The most important input features are \met\ and the two tau $\pt$s. In general,
\met\ is an important signature for models with an uncharged (N)LSP at
the end of decay chains, as these do not interact with detectors. Hence,
it is as expected to see \met\ correlating strongly with the target
class for this particular analysis, and the observation that the \ac{ML} models have modeled this correlation structure is reassuring.

\subsection{Comparison of approaches}
In~\cref{sec:cut_analysis,sec:cut_classifier,sec:mle} we have presented three
different approaches to calculating the discovery significance in a sample with
a mix of \ac{SM} and \ac{SUSY} events. The first two approaches -- cut-and-count
analysis and cutting on the \ac{ML} classifier output -- define a selection to measure the number of events observed and to compare it to the expected number of events. A properly 
trained \ac{ML} classifier will outperform or, at
worst, perform as well as the cut-and-count approach in most situations.
From the point of view of a statistical analysis it does not
matter what method the number of events is coming from. Therefore, we only compare the \ac{ML}-classifier-based cut method with the
likelihood fit of the classifier shapes.

The unbinned maximum likelihood fit described in~\cref{sec:mle} attempts to
represent the shape of the classifier output as the sum of signal (Point 12)
and background (\ac{SM}) templates.  In a fit to any other parameter point, the
\ac{SM} contribution is described purely by the background template (except for
statistical fluctuations) while the signal contribution is represented as some
combination of the signal (Point 12) and background (\ac{SM}) templates. It
follows that the true mixture parameter is the upper bound for the mixture
parameter as determined by the fit, only achievable when the signal point
considered has a completely non-\ac{SM}-like classifier shape. Since the
background yield is left unconstrained, this makes the template fit approach
conservative in the sense of discovery and will not overestimate the signal
contribution as long as the templates are chosen properly, specifically that the ratio between the two is a strictly monotonic function, see e.g.~\cite[sec.~5]{kvellestad2018signal} or \cite[fig.~1b]{cranmer2015approximating}.

The maximum likelihood fit provides the discovery significance due to Wilks'
theorem. To obtain the exclusion significance one would run the fit under
background-only assumption, i.e., fitting with just the background (\ac{SM})
template. 

Comparing the cut on the ML classifier output
with the template likelihood
fit approach, there are several factors to consider: robustness, significance
and explainability. Since the two methods are based on the same \ac{ML}
classifier, the explainability question raised by both is the same. Judging
from the selection of signal points shown in \cref{tab:point_fits}, the
template likelihood fit often leads to higher significance values while still
being a conservative estimate of the mixture parameter. An additional benefit
of the template fit method is that it makes no assumptions about the \ac{SM}
yields, i.e., it is more robust against generator acceptance effects. The
robustness of either of the two methods depends differently on the shapes of
the \ac{ML} classifier outputs, and consequently on the kinematics of the
underlying physical models. Hence, they cannot be directly compared.  

The robustness issue affects other methods for searching new physics in a
multidimensional parameter space as well. This includes the cut-and-count
analysis, although in that case the problem is easier to tackle because it is
easier to understand how the variables (e.g., \met) used in the analysis depend
on the model parameters.  The strength of \ac{ML} methods is precisely their
ability to exploit high order and non-linear correlations between features,
which necessarily makes it harder to understand how these change when changing
model parameters and thus exacerbate the robustness issue.

Finally, one is certainly not limited to just one approach. The template fit
and the \ac{ML} classifier cut approaches are statistically independent, since
they rely purely on the shape information and the yields, respectively. This
means that a simple best-of approach is justified, where one tries both
approaches and selects the optimal one on a point-by-point basis. A proper
statistical combination of the two methods should also be possible but is
outside of the scope of this study. 

\FloatBarrier
\section{Conclusions}
We have investigated the increased detectability of a supersymmetric model
featuring a gravitino LSP and a metastable sneutrino NLSP using two different
machine learning models, XGBoost and a deep neural network. 
We have considered benchmark points from a parameter space region
where superparticles are dominantly produced by electroweak processes
and where a significant number of events with two taus and one muon can
be expected.  Thus, the supersymmetric scenario serves as a test case for other
scenarios for physics beyond the Standard Model that lead to similar signatures.

We have investigated two methods of incorporating the machine learning models
into the analysis, using a threshold on the model output as event selection and
using the model output as an observable to which we perform a template fit.  
Since we do not know which region of the parameter space nature has chosen and 
optimizing the analysis for all possible parameter points is unfeasible, we have tested the 
methods' ability to generalize to parameter points they have not been
trained on. In terms of discovery significance, template fitting generally
outperforms simple cutting on classifier output. To test the generalizability for different kinematics, we scale all points to the same signal yield, that of Point 12, and perform the fit again. The results, shown in~\cref{tab:point_fits_point12yield}, indicate that the method indeed generalizes.

We observe that for parameter points where the event kinematics are highly dissimilar to those the
classifier was trained on, the classifier typically considers the data
to be more background-like. Consequently, the template fit will in such scenarios
yield conservative estimates of the signal-to-background ratio, providing robustness
to type-I errors, i.e.\ erroneously claiming discovery.
If the method is used to set exclusion limits, 
on the other hand, this effect should be taken into consideration, for
example by also considering a cut-and-count analysis, which is independent
and could be performed in parallel.

To provide additional insight into the relationship between event kinematics
and the machine learning classifier output, we have performed a Shapley
decomposition, which to a large extent matches our intuitive reasoning. The
information in the Shapley values does not provide full transparency to the
internal representation of the machine learning models -- leaving room for
future studies -- but serves as a useful tool for investigating the correlation
structure at feature level, where comparison to human knowledge is possible. 

The code used to train the classifiers and perform the template fits will
be made available at \url{https://gitlab.com/BSML/sneutrinoML} after publication.

\section*{Acknowledgements}
We would like to thank David Grellscheid and Krzystof Rolbiecki for useful discussions and
Jan Heisig for discussions and for helping to check the experimental status of our parameter
space points in the latest version of \texttt{SModelS}.
J.K.~acknowledges support from the Visiting Professor program of the
Korea Institute for Advanced Study and particularly thanks the institute
for providing a stimulating scientific environment in challenging times.
This version of the article has been accepted for publication after peer
review but is not the Version of Record and does not reflect
post-acceptance improvements, or any corrections. The Version of Record
is available online at
\url{http://dx.doi.org/10.1140/epjc/s10052-023-11532-9}.

\bibliographystyle{utphys}
\bibliography{biblio}

\appendix

\section{Parameter space points} \label{app:points}
The ten parameter space points considered in the analysis are defined by
the mass parameters listed in \cref{tab:parampoints_details}, together
with $\tan\beta=10$ and $\mu>0$. The choice of $\tan\beta=10$
yields the largest parameter space region with a sneutrino \ac{NLSP}. These values are given at the scale $Q_\text{in}=\SI{1467}{GeV}$ for Point 0 and $Q_\text{in}=\SI{1000}{GeV}$ for the other points.  Based on these inputs, the SUSY mass spectrum is calculated by \texttt{SPheno 4.0.3} and \texttt{FeynHiggs 2.14.2}.  As SM input parameters, $\alpha_s(M_Z)=0.1181$ and $m_t=\SI{173.2}{GeV}$ are used; all other \ac{SM} parameters are kept at their SPheno default values.

\begin{landscape}
    \begin{table}[tb]
        \centering 
        \caption{\label{tab:parampoints_details}Input parameter values
        for the points used in the analysis. All masses are given in
        \unit{GeV}. For expected yields, see~\cref{tab:parampoints_yields}.
        }
        \begin{tabular}{lcccccccc}
            Name & $M_{1}$ & $M_{2}$ & $M_{3}$ & $m^2_{H_d}$ & $m^2_{H_u}$ & $A_{t}$ & $A_{b}$ & $A_{\tau}$ \\
            \midrule
            Point 0  & $521.6$ & $951.3$ & $2554.5$ & $1.96 \cdot 10^7$ & $-3.66 \cdot 10^6$ & $-2.04 \cdot 10^{-3}$ & $-5544.2$ & $-3636.7$ \\
            Point 12 & $754.0$ & $267.5$ & $2466.8$ & $2.16 \cdot 10^7$ & $-2.05 \cdot 10^6$ & $-5218.4$ & $-4673.7$ & $-2933.2$ \\
            Point 13 & $903.8$ & $307.0$ & $2892.0$ & $2.09 \cdot 10^7$ & $-2.49 \cdot 10^6$ & $-5466.6$ & $-4563.7$ & $-2547.4$ \\
            Point 14 & $309.1$ & $248.2$ & $3177.3$ & $2.23 \cdot 10^7$ & $-9.47 \cdot 10^6$ & $-5503.4$ & $-5089.1$ & $-2449.2$ \\
            Point 15 & $797.8$ & $252.1$ & $3693.5$ & $2.01 \cdot 10^7$ & $-3.97 \cdot 10^6$ & $-6218.3$ & $-6009.8$ & $-2839.2$ \\
            Point 16 & $308.5$ & $258.0$ & $3722.3$ & $2.14 \cdot 10^7$ & $-9.37 \cdot 10^6$ & $-5053.8$ & $-5159.8$ & $-2526.9$ \\
            Point 20 & $599.9$ & $1000.0$ & $2724.9$ & $2.23  \cdot 10^7$ & $-3.54 \cdot 10^6$ & $-4810.6$ & $-6746.1$ & $-3744.5$ \\
            Point 30 & $446.8$ & $600.0$ & $3529.9$ & $2.02 \cdot 10^7$ & $-3.84 \cdot 10^6$ & $-5133.6$ & $-5611.5$ & $-2618.9$ \\
            Point 40 & $853.4$ & $2500.0$ & $2883.4$ & $2.50 \cdot 10^7$ & $-3.42 \cdot 10^6$ & $-5211.0$ & $-4722.5$ & $-3031.5$ \\
            Point 50 & $499.3$ & $2000.0$ & $2514.4$ & $2.40 \cdot 10^7$ & $-3.10 \cdot 10^6$ & $-2725.2$ & $-4267.0$ & $-3017.9$ \\
            \bottomrule
            & & & & & & & & \\ 
        \end{tabular}
        \begin{tabular}{lccccccccccccccc}
            Name & $M_{L11}$ & $M_{L22}$ & $M_{L33}$ &
            $M_{E11}$ & $M_{E22}$ & $M_{E33}$ & $M_{Q11}$ & $M_{Q22}$ & $M_{Q33}$ &
            $M_{U11}$ & $M_{U22}$ & $M_{U33}$ & $M_{D11}$ & $M_{D22}$ & $M_{D33}$
            \\
            \midrule
            Point 0  & $314.7$ & $314.2$ & $240.2$ & $1094.1$ & $1093.8$ & $2485.4$ & $2331.5$ & $2331.5$ & $7704.0$ & $2044.0$ & $2044.0$ & $8000.0$ & $2265.9$ & $2265.9$ & $8000.0$ \\
            Point 12  & $241.8$ & $264.2$ & $132.1$ & $1188.9$ & $1380.4$ & $1920.8$ & $2414.0$ & $2997.7$ & $4838.1$ & $ 1672.0$ & $2064.6$ & $1912.4$ & $1771.8$ & $2223.0$ & $3528.1$ \\
            Point 13  & $259.6$ & $247.9$ & $123.9$ & $1842.2$ & $1161.8$ & $1480.7$ & $2523.7$ & $2594.1$ & $4698.2$ & $1431.4$ & $1781.4$ & $2977.6$ & $2763.3$ & $2242.7$ & $3962.5$ \\
            Point 14  & $311.5$ & $236.4$ & $118.2$ & $1683.8$ & $1450.7$ & $1558.7$ & $2987.7$ & $2726.2$ & $4769.3$ & $2133.6$ & $1411.6$ & $1474.7$ & $2691.8$ & $2638.9$ & $3337.6$ \\
            Point 15  & $225.6$ & $215.6$ & $107.8$ & $1055.6$ & $1773.9$ & $1422.6$ & $2865.8$ & $2498.3$ & $4095.5$ & $1313.2$ & $1949.7$ & $1874.1$ & $2821.6$ & $1871.9$ & $3338.4$ \\
            Point 16 \\
            Point 20  & $303.4$ & $272.3$ & $136.1$ & $1767.7$ & $1702.3$ & $1133.5$ & $2722.1$ & $2696.2$ & $5449.7$ & $1809.8$ & $2127.0$ & $6552.4$ & $1632.0$ & $1605.7$ & $7612.8$ \\
            Point 30  & $384.3$ & $306.4$ & $153.2$ & $1820.0$ & $1423.8$ & $1971.9$ & $2950.8$ & $2845.8$ & $6017.0$ & $2707.6$ & $2191.0$ & $6283.7$ & $1844.6$ & $2493.7$ & $6981.8$ \\
            Point 40  & $218.3$ & $222.7$ & $111.3$ & $1115.7$ & $1300.5$ & $1553.6$ & $2497.3$ & $2493.9$ & $6005.5$ & $1327.5$ & $1072.2$ & $5567.1$ & $1526.0$ & $1037.2$ & $6749.6$ \\
            Point 50  & $210.3$ & $215.5$ & $107.7$ & $1861.1$ & $1315.2$ & $1556.5$ & $2474.9$ & $2973.2$ & $5785.7$ & $2328.9$ & $1355.2$ & $7303.1$ & $2902.0$ & $2917.2$ & $9256.5$ \\
            \bottomrule
        \end{tabular}
    \end{table}
\end{landscape}

\begin{table}[H]
	\caption{\label{tab:point0_brs}Branching ratios (BRs) of Point 0. Only BRs larger than
$1\%$ are included.}

\begin{tabular}{lcc}
\toprule
$\chione \rightarrow$ & $\stauone^* \tau^{-}$ + c.c. & $\nustau \bar{\nu}_{\tau}$ + c.c.\\
\midrule
BR$[\%]$ & $14.6 \times 2 $& $35.4 \times 2$ \\
\end{tabular}

\begin{tabular}{lcc}
\toprule
$\chitwo \rightarrow$ & $\stauone^* \tau^{-}$ + c.c. & $\nustau \bar{\nu}_{\tau}$ + c.c.\\
\midrule
BR$[\%]$ & $43.1 \times 2 $& $6.3 \times 2$ \\
\bottomrule
\end{tabular}

\begin{tabular}{lcc}
$\chionep \rightarrow$ & $\stauone^*  \nu_{\tau} $ & $\tau^+ \nustau $\\
\midrule
BR$[\%]$ & $5.7$ & $92.9$ \\
\end{tabular}

\begin{tabular}{lcccc}
\toprule
$\sem \rightarrow$  & $e^{-} \chione$ & $\chionem \nu_{e}$ & $u d \nuse$ & $c s \nuse$ \\
\midrule
BR$[\%]$ & $95.1$ & $1.8$ & $1.6$ & $1.5$ \\
\bottomrule
\end{tabular}

\begin{tabular}{lccc}
$\smum \rightarrow$ & $\mu^{-} \chione$ &$u d \nusmu$  &$c s \nusmu$  \\
\midrule
BR$[\%]$ & $96.6$ & $1.6$ & $1.4$ \\
\end{tabular}

\begin{tabular}{lccccc}
\toprule
$\stauone^- \rightarrow$ &$\nustau e^- \bar\nu_e$  & $\nustau \mu^- \bar\nu_\mu$ & $\nustau \tau^- \bar\nu_\tau$ & $\nustau d \bar{u}$  & $\nustau s \bar{c}$ \\
\midrule
BR$[\%]$ & $11.6$ & $11.6$ & $10.0$ &$34.7$ &$32.1$ \\
\bottomrule
\end{tabular}

\begin{tabular}{lcc}
$\nuse \rightarrow$ & $\chione \nu_e$ & $\nustau e^- \tau^+$ \\
\midrule
BR$[\%]$ & $96.0$ & $4.0$\\
\end{tabular}

\begin{tabular}{lcc}
\toprule
$\nusmu \rightarrow$ & $\chione \nu_{\mu}$ & $\nustau \mu^- \tau^+$ \\
\midrule
BR$[\%]$ & $52.8$ & $47.2$ \\
\bottomrule
\end{tabular}
\end{table}

\begin{table}[H]
    \caption{\label{tab:point12_brs}Branching ratios (BRs) of Point 12. Only
    BRs larger than $1\%$ are included.}

\begin{tabular}{lcccccc}
\toprule
$\chione \rightarrow$ & $\stauone^* \tau^{-}$ + c.c. & $\nustau \bar{\nu}_{\tau}$ + c.c. & $\se^* e^-$ + c.c. & $\nuse \nu_e $ + c.c. & $\smu^* \mu^-$ + c.c. & $\nusmu \nu_{\mu}$ + c.c.\\
\midrule
BR$[\%]$ & $17.6 \times 2$ & $21.3 \times 2$ & $3.2 \times 2$ & $4.9 \times 2$ & $1.1 \times 2$ & $2.1 \times 2$ \\
\bottomrule
\end{tabular}

\begin{tabular}{lccccc}
$\chionep \rightarrow$ & $\stauone^*  \nu_{\tau} $ & $\tau^+ \nustau $ & $e^+ \nuse$ & $\smu^* \nu_{\mu}$ & $\mu^+ \nusmu$\\
\midrule
BR$[\%]$ & $37.3$ & $45.7$ & $10.3$ & $2.3$ & $4.4$ \\
\end{tabular}

\begin{tabular}{lccccc}
\toprule
$\sem \rightarrow$ & $\stauone \tau^+ e^-$ & $\stauone^* \tau^- e^- $ & $\nustau e^- \bar{\nu}_{\tau}$ & $\nustau^* e^- \nu_{\tau}$ & $\nustau^* \tau^- \nu_e$ \\
\midrule
BR$[\%]$ & $3.0$ & $8.1$ & $5.7$ & $14.5$ & $68.4$ \\
\end{tabular}

\begin{tabular}{lccccc}
\toprule
$\smum \rightarrow$ & $\stauone \tau^+ \mu^-$ & $\stauone^* \tau^- \mu^- $ & $\nustau \mu^- \bar{\nu}_{\tau}$ & $\nustau^* \mu^- \nu_{\tau}$ & $\nustau^* \tau^- \nu_{\mu}$\\
\midrule
BR$[\%]$ & $4.3$ & $9.5$ & $6.7$ & $14.1$ & $65.0$\\
\end{tabular}

\begin{tabular}{lccccc}
\toprule
$\stauone^- \rightarrow$ &$\nustau e^- \bar\nu_e$  & $\nustau \mu^- \bar\nu_\mu$ & $\nustau \tau^- \bar\nu_\tau$ & $\nustau d \bar{u}$  & $\nustau s \bar{c}$ \\
\midrule
BR$[\%]$ & $11.3$ & $11.3$ & $10.1$ & $33.8$ & $33.4$ \\
\end{tabular}

\begin{tabular}{lcccccc}
\toprule
$\nuse \rightarrow$ & $\nustau e^- \tau^+$ & $\nustau \nu_e \bar{\nu}_{\tau}$ & $\nustau^* \nu_e \nu_{\tau}$ & $\stauone \tau^+ \nu_e$ & $\stauone^* e^- \nu_{\tau}$ & $\stauone^* \tau^- \nu_e$ \\
\midrule
BR$[\%]$ & $27.7$ & $7.3$ & $25.8$ & $2.6$ & $30.1$ & $6.4$ \\
\end{tabular}

\begin{tabular}{lcccccc}
\toprule
$\nusmu \rightarrow$ & $\nustau \mu^- \tau^+$ & $\nustau \nu_{\mu} \bar{\nu}_{\tau}$ & $\nustau^* \nu_{\mu} \nu_{\tau}$ & $\stauone \tau^+ \nu_{\mu}$ & $\stauone^* \mu^- \nu_{\tau}$ & $\stauone^* \tau^- \nu_{\mu}$ \\
\midrule
BR$[\%]$ & $28.1$ & $7.4$ & $20.9$ & $3.5$ & $32.9$ & $7.1$ \\
\bottomrule
\end{tabular}

\end{table}
\FloatBarrier

\pagebreak
\section{Results with equal signal yields}
Here, \cref{tab:point_fits} is recalculated with test data scaled to the expected signal yields of Point 12, i.e., the true mixture parameter is equal for all points. The signal events are distributed between the different production channels as before, i.e, according to the cross sections.
\begin{table}[ht]
    \centering
    \small
    \caption{\label{tab:point_fits_point12yield}
    Template fit results for all parameter points, scaled to the same yields as Point 12, but with each point's respective kinematics. All points have $\alpha_{\text{true}}=0.07$, and both classifiers used are trained on Point 12.}
    \begin{tabular}{lSllSSSS}
        & $\alpha_{\mathrm{true}}$ & $\hat{\alpha}_{\text{XGB}}$  &
        $\hat{\alpha}_{\text{DNN}}$ & $z_{\text{XGB}}$ & $z_{\text{DNN}}$
        & $z_{\text{XGB-cut}}$ & $z_{\text{DNN-cut}}$ \\
        \midrule
        Point 12 & 0.070  & $0.069 \pm 0.011$   & $0.071 \pm 0.011$ & 15.7 & 18.5 & 15.5 & 16.7  \\
        Point 13 & 0.070  & $0.060 \pm 0.011$   & $0.063 \pm 0.011$ & 13.9 & 16.8 & 13.7 & 14.6  \\
        Point 14 & 0.070  & $0.058 \pm 0.011$   & $0.059 \pm 0.011$ & 13.2 & 15.5 & 12.4 & 13.0  \\
        Point 15 & 0.070  & $0.055 \pm 0.011$   & $0.055 \pm 0.01$  & 12.4 & 14.4 & 10.6 & 11.1  \\
        Point 16 & 0.070  & $0.046 \pm 0.012$   & $0.045 \pm 0.01$  & 10.9 & 12.2 & 10.3 & 10.5  \\
        Point 20 & 0.070  & $0.061 \pm 0.011$   & $0.063 \pm 0.011$ & 14.4 & 17.0 & 14.1 & 15.3  \\
        Point 30 & 0.070  & $0.083 \pm 0.012$   & $0.083 \pm 0.012$ & 19.3 & 22.4 & 17.4 & 18.7  \\
        Point 40 & 0.070  & $0.038 \pm 0.01$    & $0.040 \pm 0.009$ & 8.9  & 10.5 & 10.0 & 10.4  \\
        Point 50 & 0.070  & $0.052 \pm 0.01$    & $0.051 \pm 0.01$  & 12.3 & 13.4 & 11.1 & 11.6  \\
        Point 0  & 0.070  & $0.014 \pm 0.008$   & $0.012 \pm 0.008$ & 3.3  & 3.2  & 4.4  &  3.5  \\
        \bottomrule
    \end{tabular}
\end{table}
\FloatBarrier

\section{Effect of theoretical uncertainty on the classifier outputs}
\label{sec:signal_variation_comparison}
\begin{figure}
    \centering
    \begin{subfigure}[t]{0.47\textwidth}
        \includegraphics[width=\textwidth]{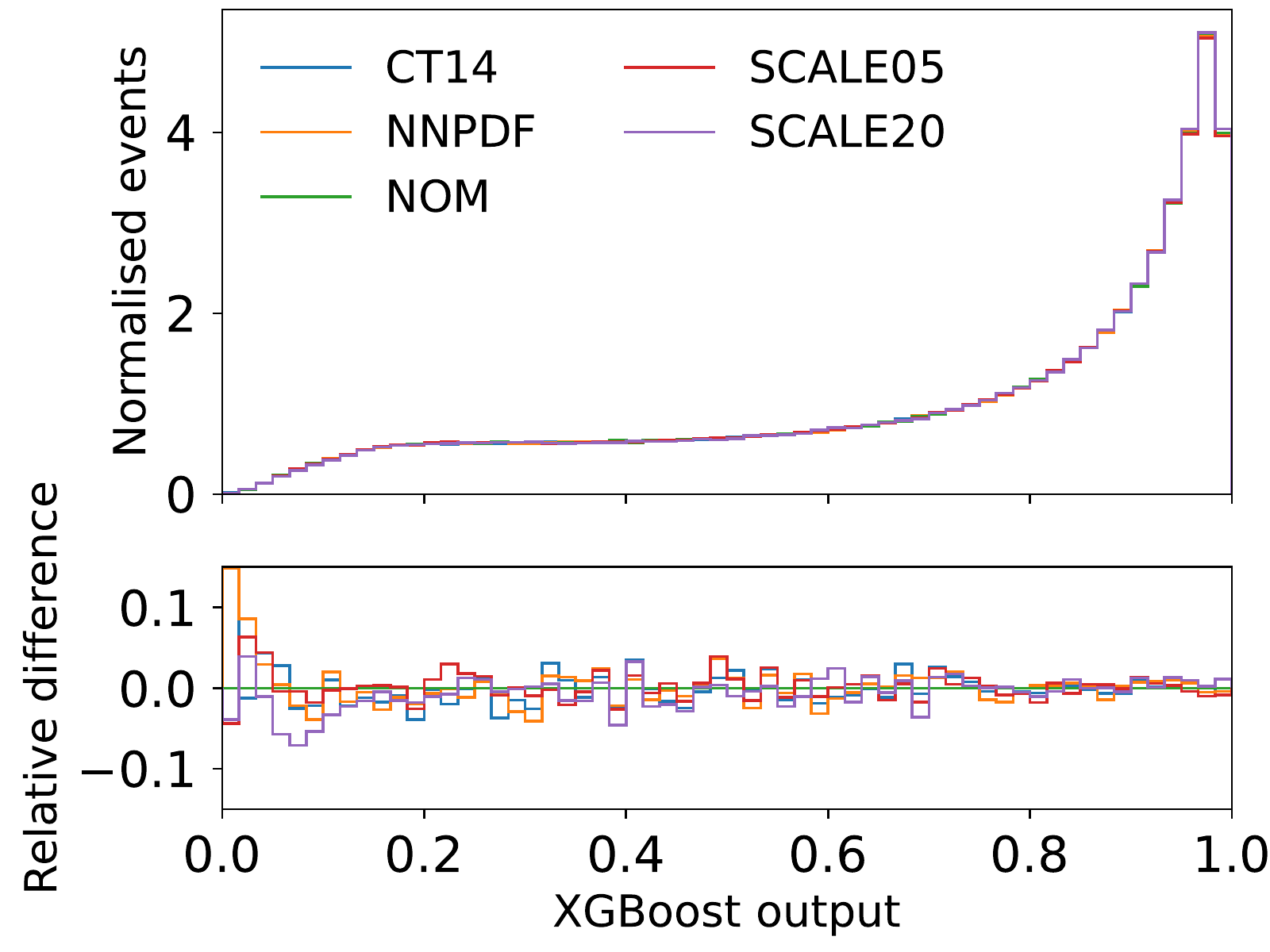}\vspace{-1.5ex}
        \caption{\label{fig:signal_variation_point12_xgb}}
    \end{subfigure}
    \begin{subfigure}[t]{0.47\textwidth}
        \includegraphics[width=\textwidth]{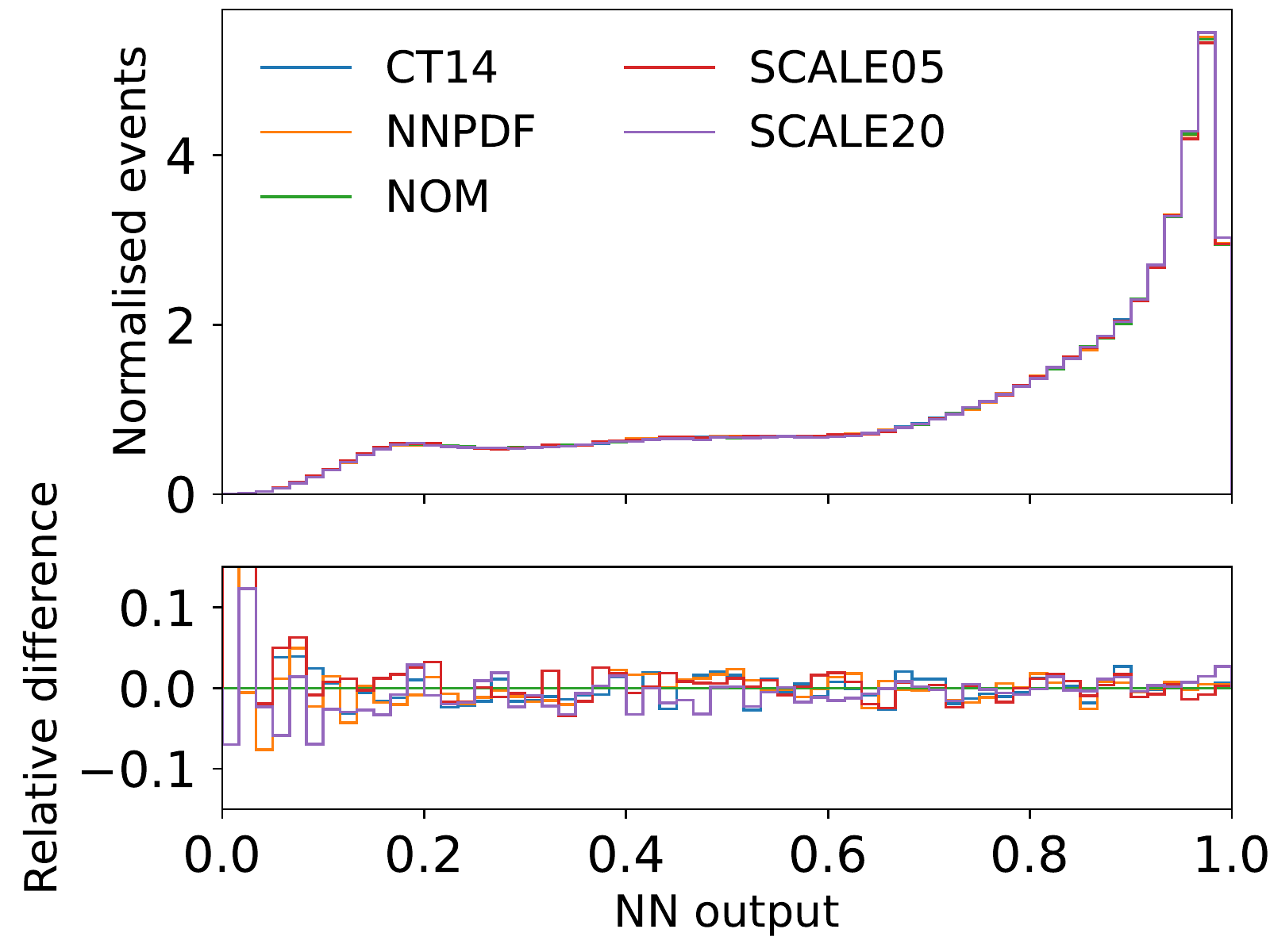}\vspace{-1.5ex}
        \caption{\label{fig:signal_variation_point12_nn}}
    \end{subfigure}
    \caption{\label{fig:signal_variation_point12}Output distributions from (\protect\subref{fig:signal_variation_point12_xgb}) the XGBoost classifier and (\protect\subref{fig:signal_variation_point12_nn}) the DNN classifier for nominal and alternative signal samples for Point 12. The lower panel shows the relative difference between each alternative PDF setting and the nominal one.}
\end{figure}

\begin{figure}
    \centering
    \begin{subfigure}[t]{0.47\textwidth}
        \includegraphics[width=\textwidth]{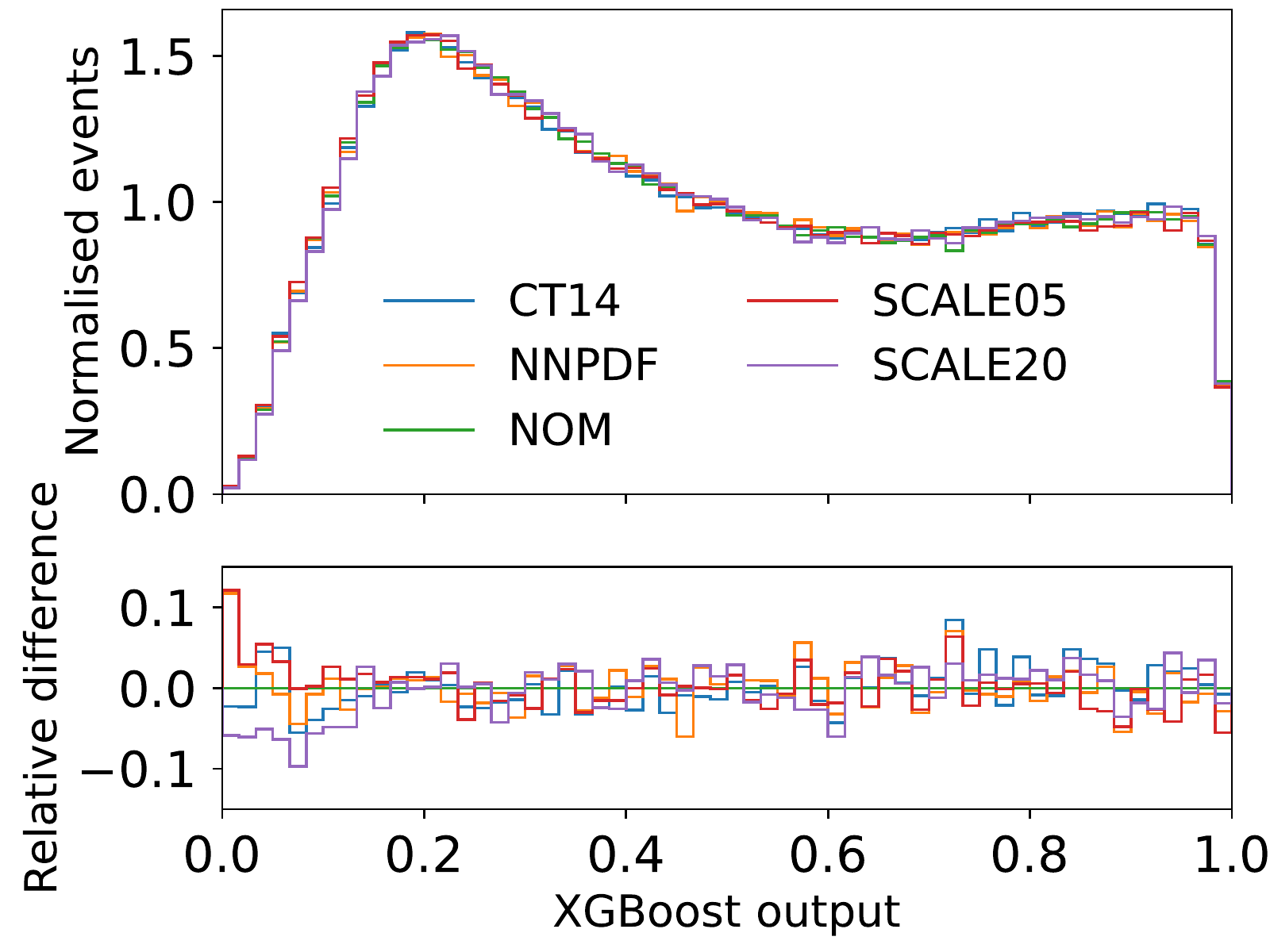}\vspace{-1.5ex}
        \caption{\label{fig:signal_variation_point0_xgb}}
    \end{subfigure}
    \begin{subfigure}[t]{0.47\textwidth}
        \includegraphics[width=\textwidth]{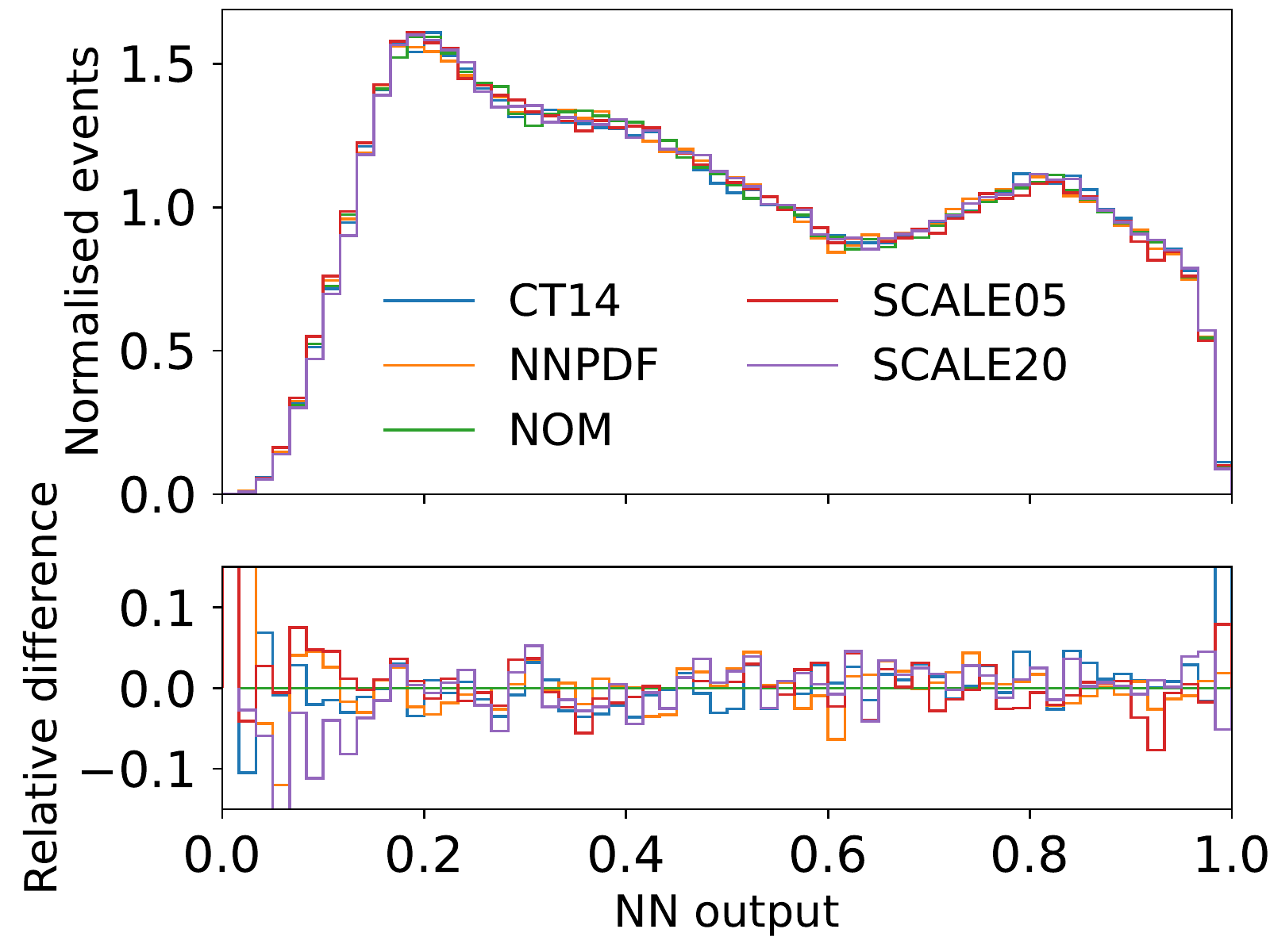}\vspace{-1.5ex}
        \caption{\label{fig:signal_variation_point0_nn}}
    \end{subfigure}
    \caption{\label{fig:signal_variation_point0}Output distributions from (\protect\subref{fig:signal_variation_point0_xgb}) the XGBoost classifier and (\protect\subref{fig:signal_variation_point0_nn}) the DNN classifier for nominal and alternative signal samples for Point 0. The lower panel shows the relative difference between each alternative PDF setting and the nominal one.}
\end{figure}

In order to evaluate the theoretical uncertainty originating from the choice of
PDF set in the signal generation, classifier outputs are computed for signal samples produced using
the alternative PDF sets described in \cref{sec:evtgen}. These are shown in \cref{fig:signal_variation_point12}
for Point~12 and in \cref{fig:signal_variation_point0} for Point~0. In both cases, the classifiers trained on nominal Point~12 samples are used. The lower panels in the plots show the ratio of each alternative sample to the nominal. Both classifiers are very robust to the kinematic differences between the alternative samples, showing no apparent bias across the output range.
Hence, the results of the maximum likelihood fit to the classifier
outputs are expected to be dominated by the statistical uncertainty, which was also verified by testing.


\end{document}